\documentclass[11pt,a4paper]{article}
\usepackage{jheppub_2}
\usepackage{amsmath}
\usepackage{amssymb}
\usepackage{graphicx}
\usepackage{color}
\usepackage[normalem]{ulem}
\usepackage[utf8]{inputenc}
\usepackage{mathrsfs}
\usepackage{xfrac}
\usepackage{color}
\usepackage{cleveref}
\usepackage{caption}
\usepackage{subcaption}
\usepackage{feynmp}
\usepackage{xcolor}

\newcommand{\MeV}{\text{MeV}}
\newcommand{\GeV}{\text{GeV}}
\newcommand{\TeV}{\text{TeV}}
\newcommand{\MPl}{M_{\rm Pl}}
\newcommand{\CP}{\emph{CP}~}
\newcommand{\p}{\varphi}
\newcommand{\Lag}{\mathcal{L}}
\newcommand{\dgy}{c_{g_Y}}
\newcommand{\dgw}{c_{g_2}}

\newcommand{\deltaW}{\delta \alpha_2}
\newcommand{\eff}{\text{eff}}
\newcommand{\SM}{\text{SM}}
\newcommand{\EW}{\text{EW}}
\newcommand{\beq}{\begin{equation}}
\newcommand{\eeq}{\end{equation}}

\newcommand{\Hcut}{\Lambda_6/\sqrt{c_6}}

\preprint{\\ UCI-HEP-TR-2019-13}

\title{Electroweak Baryogenesis from Temperature-Varying Couplings}
\author[a]{Sebastian~A.~R.~Ellis}
\author[b]{Seyda Ipek}
\author[c]{and Graham White}

\affiliation[a]{SLAC National Accelerator Laboratory, 2575 Sand Hill Road, Menlo Park, CA 94025, USA}
\affiliation[b]{Department of Physics and Astronomy, University of
California, Irvine, CA 92697-4575 USA}
\affiliation[c]{TRIUMF Theory Group, 4004 Wesbrook Mall, Vancouver, B.C. V6T2A3, Canada}

\emailAdd{sarellis@slac.stanford.edu}
\emailAdd{sipek@uci.edu}
\emailAdd{gwhite@triumf.ca}
\date{\today}
\abstract{
The fundamental couplings of the Standard Model are known to vary as a function of energy scale through the Renormalisation Group (RG), and have been measured at the electroweak scale at colliders. However, the variation of the couplings as a function of temperature need not be the same, raising the possibility that couplings in the early universe were not at the values predicted by RG evolution. We study how such temperature-variance of fundamental couplings can aid the production of a baryon asymmetry in the universe through electroweak baryogenesis. We do so in the context of the Standard Model augmented by higher-dimensional operators up to dimension 6. 
}

\begin{document}
\maketitle

\section{Introduction}
The Standard Model (SM) of particle physics provides an elegant explanation for many of the observable phenomena in the universe. However, it fails to explain a few critical phenomena, including one which is crucial to our existence, namely the origin of the matter-antimatter asymmetry in the universe. Through cosmological observations from the Cosmic Microwave Background (CMB)~\cite{Aghanim:2018eyx} and Big Bang Nucleosynthesis (BBN)~\cite{Riemer-Sorensen:2017vxj}, this asymmetry is measured to be 
\begin{equation}
   Y_B = \frac{n_B}{s} \sim 10^{-10} \ .
\end{equation}
The SM fails to explain the origin of the baryon asymmetry of the universe (BAU) because it does not satisfy two of the three \emph{Sakharov conditions}~\cite{Sakharov:1967dj}. While the SM has $SU(2)_L$ sphalerons which conserve $B-L$ but violate $B+L$, it does not have enough $CP$-violation, nor does it contain the necessary out-of-equilibrium process. This failure of the otherwise extremely successful SM is a compelling reason to \emph{demand} new physics beyond the Standard Model (BSM). 

BSM models that address the problem of generating the BAU often include new fields that couple to the Higgs field such that the electroweak (EW) transition becomes a first-order phase transition, which provides the out-of-equilibrium condition. In addition, these new couplings can be the source of the additional $CP$ violation needed to explain the BAU. (For recent reviews, see \cite{Morrissey:2012db,White:2016nbo}.)

Both the $CP$ violation and the out-of-equilibrium processes involved in the generation of the BAU depend on the gauge and Yukawa couplings at the time of the EW transition. For example, it has been shown that by modifying Yukawa couplings in the early universe,  $CP$ violation in the SM can be enhanced \cite{Berkooz:2004kx, Baldes:2016gaf, Baldes:2016rqn, vonHarling:2016vhf, Bruggisser:2017lhc}. It has also been shown that gauge couplings in the early universe can be different from their expected values given renormalisation group running in the SM \cite{Ipek:2018lhm}. 

In this work we investigate how changes in weak and strong coupling constants during the EW transition affect EW baryogenesis scenarios. To be more specific, we use an Effective Field Theory (EFT) framework which encapsulates the interactions of light fields with decoupled new physics. This formalism allows us to introduce the operators required to satisfy the Sakharov conditions, as well as modify the SM gauge couplings. The effect of operators required to ensure a first order phase transition and provide new \CP violation are well understood, and have been studied extensively in the literature \cite{Grojean:2004xa,Delaunay:2007wb,deVries:2017ncy,Balazs:2016yvi,Chala:2018ari,deVries:2018tgs}. The novel aspect of our work is to include dimension-5 operators involving new scalar fields, which change the effective gauge couplings in the early universe.

The effect of modified gauge couplings can be separated into two categories. The most intuitive category is how changing the $SU(2)_L$ and $SU(3)_c$ coupling constants, $\alpha_2$ and $\alpha_3$, directly affect the rates of the EW and strong sphalerons respectively. Since sphaleron rates for an $SU(N)$ symmetry group scale as $\alpha_N^5$, it is clear that even modest variations in the coupling constants can result in large variations in the rates. Since EW baryogenesis relies on out-of-equilibrium $SU(2)_L$ sphaleron processes to generate a baryon asymmetry, one would expect that making these processes more efficient should enhance the production of baryons, as long as the same enhanced sphalerons do not subsequently erase the produced asymmetry. Additionally, EW sphaleron transitions are more effective in the presence of a chiral asymmetry (sometimes called a \CP asymmetry), generated by interactions on the wall of the expanding bubbles of true vacuum. This chiral asymmetry is efficiently washed out by strong sphalerons, so that reducing their strength should also help with generating the observed baryon asymmetry. The less intuitive category of effects is how changing gauge couplings affects the finite-temperature Higgs potential, and the calculation of the transport of particles in the vicinity of the true vacuum bubble wall. These effects are mostly due to modified thermal masses of the relevant gauge bosons and altered interaction rates. 

We perform a calculation of these effects on the baryon asymmetry, as well as present models which could give rise to the variation in the gauge couplings. We do so in a model which would otherwise be ruled out by current electron electric dipole moment (EDM) constraints, demonstrating the power of such variations in the gauge couplings. 

In  \Cref{sec:model} we present the EFT model we will use for our analysis. In  \Cref{sec:EWPT} we discuss the EW phase transition, and the possible effects of modifying gauge couplings. In \Cref{sec:BAU} we discuss in greater detail how we compute the baryon asymmetry, starting from the requirement of new sources of \CP violation, continuing with a discussion of how the modified gauge couplings affect the computation of the obtained baryon asymmetry, and ending with a discussion of how washout from enhanced $SU(2)_L$ sphalerons in the broken phase of EW symmetry can be avoided. This section, and in particular  \Cref{fig:EWBG,fig:EWBGalpha3}, constitute the main results of our paper. In \Cref{sec:varyingcouplings} we propose models which could give rise to the required modifications of gauge couplings in the early universe, which are consistent with current measurements. Finally, in \Cref{sec:conclusions} we present our concluding remarks.

\section{Augmenting the Standard Model: Effective Operators}
\label{sec:model}
In this section we describe the general features of the model(s) we work with. 
The failure of the SM to reproduce the observed baryon asymmetry \emph{requires} the presence of new fields and interactions. When there is a separation of scales between heavy and light fields, the former can be integrated out. The effect of the heavy fields is then encapsulated in the Wilson coefficients of higher dimensional operators involving only the remaining light fields, suppressed by the heavy field scale. The SM augmented by such higher dimensional operators involving only SM fields is commonly referred to as the SM Effective Field Theory (SMEFT). In our analysis we will invoke additional fields that are near or below the EW scale, so that it is a more general EFT framework. 
Current constraints on the scale of new physics from the LHC can be enough to warrant the use of an EFT approach to study the effects of decoupled new physics. 
Additionally, in the case of understanding mechanisms of baryogenesis, the use of an EFT is further validated by the strong constraints on new \CP violation beyond the SM, coming from the non-observation of neutron and electron EDMs.\footnote{There are also strong constraints on new sources of \CP violation from quark flavour observables (see e.g. \cite{Altmannshofer:2013lfa, Isidori:2013ez}), but these involve flavour-changing operators, which we do not consider here.} Together these can constrain new physics up to the PeV scale in some models.

In this analysis, we consider a minimal set of higher dimensional operators that need to be added to the SM to ensure successful baryogenesis in the context of varying $\alpha_2$ and $\alpha_3$ in the early universe. 

The EW phase transition in the SM is a crossover, meaning that the Sakharov condition of out-of-equilibrium processes is not satisfied. It has been shown that a dimension-6 Higgs operator is sufficient to turn this crossover into a first-order phase transition, thereby satisfying  one of the Sakharov conditions \cite{Grojean:2004xa,Chala:2018ari,Delaunay:2007wb}. Therefore, \textbf{(1)} we include the operator $\mathcal{O}_6 = |H|^6$, where $H$ is the $SU(2)_L$ Higgs doublet field, which contains the physical Higgs field $h$.

The SM has $CP$ violation in the quark sector, but it is much too small to ensure successful baryogenesis \cite{Huet:1994jb}. To compensate for this, in SMEFT, new $CP$ violation through effective Higgs-fermion operators are added. It is expected that the top-Higgs operator is the most important due to the large top Yukawa. However, it has been shown that constraints on this term rule it out for generating the BAU \cite{deVries:2017ncy}. We will show later that this operator can generate enough $CP$ violation while being consistent with EDM measurements when we allow for gauge couplings to vary during the EW transition. Hence \textbf{(2)} we add the top-Higgs operator, $\mathcal{O}_{tH} = \bar{Q}_L \tilde{H} t_R |H|^2$, with $\tilde{H} = i \sigma_2 H^*$, to our Lagrangian.

Finally,  \textbf{(3)} we add dimension-5 operators parameterising interactions between new scalar singlets and the EW and strong gauge kinetic terms, which will enable the variation of $\alpha_2$ and $\alpha_3$ in the early universe through the dynamics of these scalar fields.  

Given these three additions to the SM, we may write the following Lagrangian for the model under consideration
\begin{align}
\nonumber \Lag\supset &-\frac{1}{4g_Y^2}\left(1- \frac{c_{g_Y}\p_Y}{\Lambda _Y} \right) B^{\mu \nu} B_{\mu \nu}-\frac{1}{4g_2^2}\left(1- \frac{c_{g_2}\p_2}{\Lambda _2} \right) W^{\mu \nu, a} W^a_{\mu \nu}\\ &-\frac{1}{4g_3^2} \left(1-\frac{c_{g_3}\p_3}{\Lambda_3} \right) G^{\mu \nu, A} G^A_{\mu \nu}- V_{SM}(H)- \frac{c_6}{\Lambda_6^2}|H|^6+\frac{\delta_{CPV}}{\Lambda ^2 _{\rm CPV}}\bar{Q}_L \tilde{H} t_R |H|^2,  \label{eq:model}
\end{align}
where $V_{SM}(H)=\mu_H^2|H|^2+\lambda_H|H|^4$ is the SM Higgs potential and $\Lambda_X$ are various new physics scales, presumably unrelated to each other. At first, we will be agnostic regarding how these higher dimensional operators are generated. In \Cref{sec:varyingcouplings} we propose two UV completions for the dimension-5 operators, both of which will have testable properties. 

The first three terms in \Cref{eq:model} are crucial elements of this work. These terms ensure that the effective weak and strong couplings become field-dependent, 
\begin{align}
\frac{1}{g_{i,\eff}^2}\equiv \frac{1}{g_i^2}\left(1- \frac{c_{g_i}\p_i}{\Lambda _i}\right). 
\end{align}
Therefore, if the scalar fields $\p_i$ have non-trivial dynamics in the early universe, this will lead to a variation of the gauge couplings at early times/large temperatures that does not immediately follow from the Renormalisation Group equations. We will eventually consider UV completions for these dimension-5 operators which will enable us to change the values of the gauge couplings at temperatures around the EW phase transition. The change in gauge couplings will be parameterised henceforth by
\begin{align}
 \delta \alpha_i = (\alpha_{i, \eff}-\alpha_{i, \SM})|_{T=T_{\EW}}~.
 \end{align}
 In the following sections we investigate how this change in gauge couplings affects the EW phase transition, the resulting gravitational wave spectrum, and most importantly, the generation of the observed baryon asymmetry.

\section{Varying the Weak Coupling Constant and the Electroweak Phase Transition}
\label{sec:EWPT}
The strong and weak coupling constants feed into the calculation of the BAU in EW baryogenesis (EWBG) scenarios in several different ways and it is not always straightforward to disentangle these contributions. In this section we describe how the order of the phase transition changes with varying the weak coupling from its SM value while remaining agnostic to the origin of this deviation. Note that the strong coupling constant does not play a role in the EW phase transition. However, variations in $\alpha_3$ will be important for generating the BAU, as will be discussed in the next section.

At  temperatures higher than the EW scale ($T\gtrsim100~\GeV$), the Higgs potential gets finite-temperature corrections due to the interactions of the Higgs field, $h$, with particles inside the plasma. The finite temperature Higgs potential is calculated extensively in the literature. The effective potential can be written as a sum of zero-temperature and finite-temperature pieces,
\begin{equation}
    V(h,T)= V_0(h) + V_{\rm CW} (h) + V_T(h,T) \ ,
\end{equation}
where the first term is the tree-level potential and middle term is the zero-temperature Coleman-Weinberg loop correction. Loop corrections include all particles that interact with the Higgs. However, it is enough to focus on the ones with largest couplings, \emph{i.e.} Higgs self-coupling, gauge bosons and the top quark. For a given value of $\Hcut$, the parameters in the effective potential are fixed by requiring that the zero-temperature Higgs mass and vacuum expectation value (vev) are given at 1-loop by their experimentally measured values for a renormalization scale of $\mu= m_Z$. At finite temperature the renormalization scale is set to the temperature $\mu = T$.\footnote{Note that the analysis of the thermal parameters has some renormalization dependence as shown in \cite{Kainulainen:2019kyp}. We find that this effect gets accentuated near the limits of our parameters space, i.e. $\delta \alpha _2 \sim 0.1$. Dealing with this uncertainty through dimensional reduction techniques is left to future work.} The finite temperature corrections to the tree-level potential are
\begin{equation}
    V_T =  \sum _{i \in {\rm bosons}}  n_i\frac{T^4}{2 \pi ^2} J_B\left[ \frac{m_i^2 + \Pi _i}{T^2} \right]+ \sum _{i \in {\rm fermions}}  n_i\frac{T^4}{2 \pi ^2} J_F\left[ \frac{m_i^2 }{T^2} \right] \ .
\end{equation}
In the above the $\Pi _i$ terms are Debye masses that result from a resummation and are included to prevent the breakdown of perturbation theory, see, \emph{e.g.}, \cite{Parwani:1991gq,Arnold:1992rz,Curtin:2016urg}. 

In the SM, setting $c_6=0$ in \Cref{eq:model}, the finite-temperature potential has a simple form 
\begin{align}
    V(h,T) \sim D(T^2-T_0^2) h^2-ET h^3 +\frac{\lambda}{4} h^4~, \label{eq:VTh}
\end{align}
where
\begin{align*}
    D &= \frac{2 M_W^2 +M_Z^2 +2 m_t^2}{4 v^2}~,~~~
    E = \frac{2 M_W^3+M_Z^3}{6\pi v^3}~,~~~~
    T_0^2 = \frac{m_h^2}{4D}~ ,
\end{align*}
and $v=246~\GeV$ is the zero-temperature Higgs vev. As can be seen from the above equation, the Higgs potential at high temperatures is dominated by the quadratic term and as such the EW symmetry is not broken. As the universe cools down, the potential acquires a second minimum at non-zero $\langle h\rangle$. At a critical temperature $T_c$, this minimum becomes the global minimum and the Higgs field has a non-zero vev $v_c$. 

The details of this transition are crucial for generating the BAU.  Specifically, in order to satisfy the out-of-equilibrium condition, a first-order phase transition (FOPT) is required. The strength of the phase transition is measured by comparing $v_c$ and the the size of the barrier that separates the false vacuum at $\langle h\rangle =0$ and the true vacuum at $\langle h\rangle =v_c$. In the SM and in most BSM scenarios, this condition translates into $v_c/T_c \gtrsim 1$. (In \Cref{sec:BAU} we will show how this condition is modified in our model.) From \Cref{eq:VTh}, and taking $g_2 \to g_2 +\delta g_2$ at $T_c$, we get 
\begin{equation}
    \frac{v _c}{T_c} \sim \frac{2E}{\lambda}\bigg\rvert_{T=T_c} \sim \frac{1530~\GeV^2}{m_h^2} \times \frac{(g_2+\delta g_2)^3}{g_2^3} \sim 0.1\,\frac{(g_2+\delta g_2)^3}{g_2^3}~, \label{eq:foptcondition}
\end{equation}
where we ignore variations in $g_Y$.\footnote{We numerically verified that the dependence on $g_Y$ is very weak.} It can be seen that the condition for a FOPT is not satisfied in the SM. From the above expression, it is clear that in order to get a FOPT, the EW coupling must be considerably modified.\footnote{The obtained order parameter is actually smaller than the above scaling would suggest after one includes all 1-loop corrections to the Higgs potential, including the so-called ``daisy" corrections.}  Throughout our paper we will refer to $v_c/T_c$ as the measure of the strength of the FOPT, as opposed to $v_n/T_n$, the corresponding quantities when bubbles of true vacuum begin to nucleate. Our use of $v_c/T_c$ is somewhat conservative, as it is typically required to be larger than $v_n/T_n$.

\begin{figure}[t]
\centering
\includegraphics[scale=0.45]{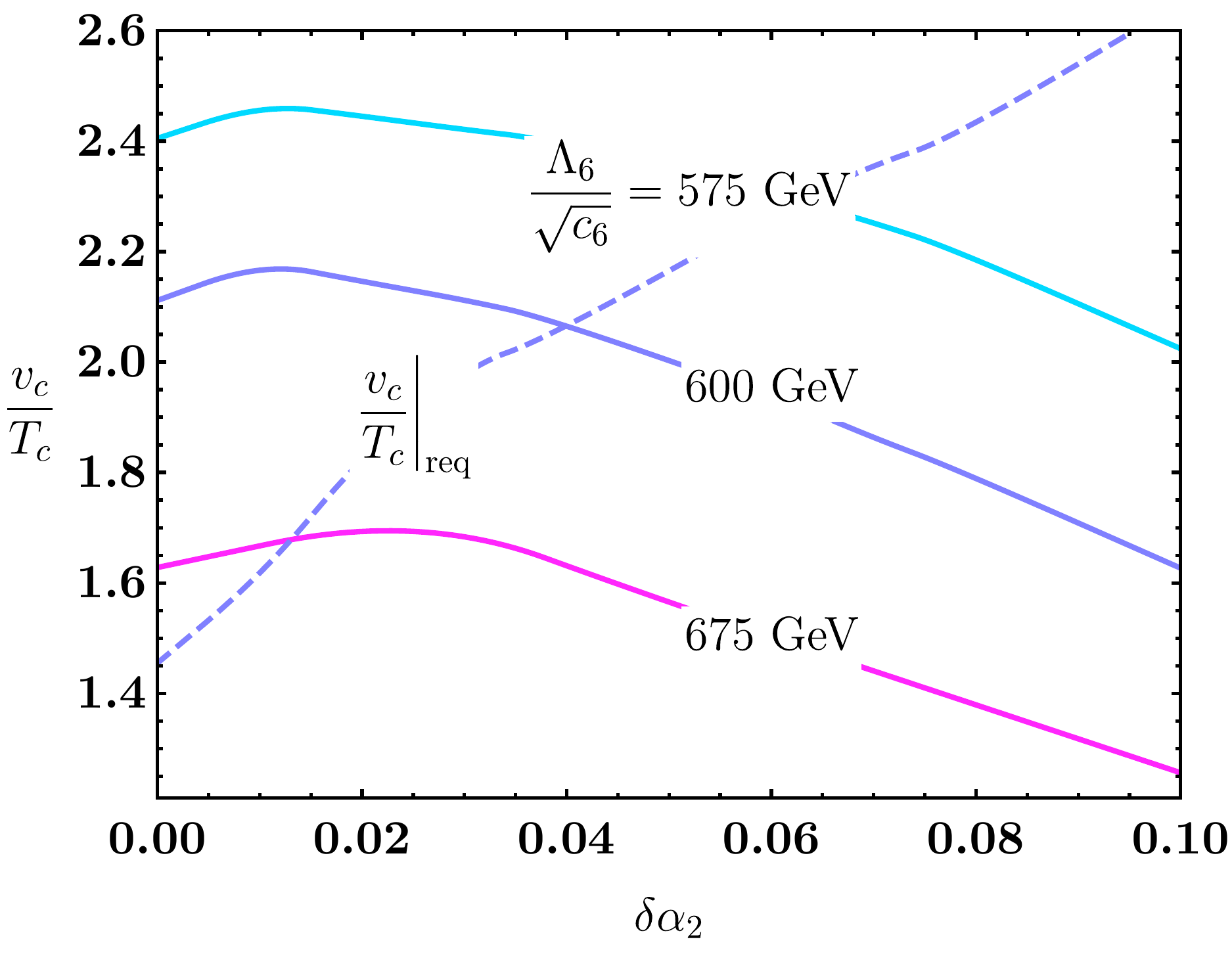}
\caption{Ratio of the Higgs vev at the critical temperature $T_c$ to the critical temperature as  a function of $\deltaW$ for different values of the dimension-6 operator, $\Hcut=575, 600, 675$~GeV. The dashed line shows the required $v_c/T_c$ for a first-order phase transition given $\Hcut=600$~GeV. Changing $\Hcut$ does not affect this dashed line substantially. For details see \Cref{washout.SEC}.}
\label{fig:vcTc}
\end{figure}

Alternatively, the barrier between the true and false vacua can be generated at tree level via a non-renormalizable $|H|^6$ operator in the potential \cite{Grojean:2004xa}, giving rise to a Higgs field 6-point interaction,
\begin{equation}
    V_0 = \frac{\mu ^2}{2} h^2 -\frac{1}{4} \lambda h^4 +\frac{c_6}{8\Lambda_6^2}  h^6~.
\end{equation}
This is what we do, as included in the Lagrangian in \Cref{eq:model}. 
In this case $v_c $ is larger than $T_c$ for $538~\GeV \lesssim \frac{\Lambda_6} {\sqrt{c_6}}\lesssim 800~\GeV$ with the lower bound set by the requirement that thermal tunneling needs to be at some stage faster than the Hubble rate \cite{Chala:2018ari}. This statement only applies in the absence of modifications to the gauge couplings. As will be shown later, this range will be altered as a function of the modified couplings.

In \Cref{fig:vcTc} we show how the order parameter $v_c/T_c$ changes while varying $\deltaW$ and different values of $\Hcut$. It can be seen that one obtains a large $v_c/T_c$ for small values of $\Hcut$, as expected. Raising the weak coupling constant raises $v_c/T_c$ for small values of $\deltaW$. Large variations in the weak coupling has the opposite effect. We suspect at larger values of $\deltaW$, contributions from daisy diagrams become more important than the leading order analysis, as in \Cref{eq:foptcondition}.

Before moving on to calculating the BAU, we also note that changing the weak coupling constant also changes the gravitational wave spectrum. expected from an EW phase transition. We include a brief discussion of this in \Cref{sec:GW}.

\section{Obtaining The Baryon Asymmetry and Avoiding Washout}
\label{sec:BAU}
In this section we calculate the baryon asymmetry produced in a SMEFT scenario with varying weak and strong coupling constants.

\subsection{Baryon Asymmetry with Modified Gauge Coupling Constants}
\begin{figure}[t]
\centering
\includegraphics[scale=0.5]{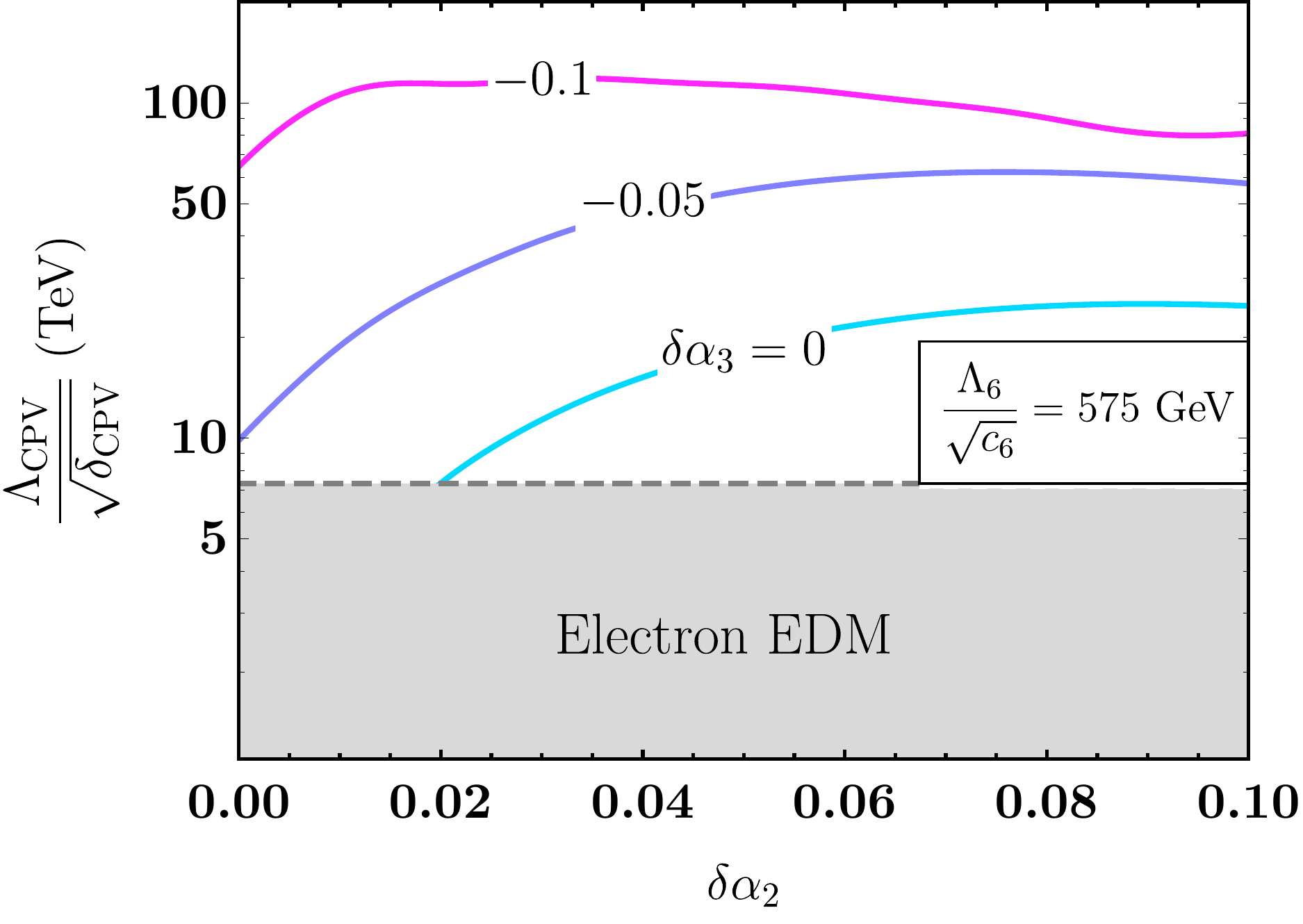}
\caption{Shown here are contours where the correct baryon asymmetry is obtained as a function of $\deltaW$ and the scale of new \CP violation $\Lambda_{\rm CPV}/\sqrt{\delta_{\rm CPV}}$, for different choices of $\delta\alpha_3$ and a fixed choice of $\Hcut = 575$ GeV. In light blue is where $\alpha_3$ is kept at its SM value, while the purple and pink contours are shown for $\delta\alpha_3= -0.05, -0.1$ respectively. The non-observation of an electron EDM requires that $\Lambda_{\rm CPV}/\sqrt{\delta_{\rm CPV}}>7.3$ TeV, shown as the grey shaded region.}
\label{CPV.FIG}
\end{figure}

\begin{figure}[t]
    \centering
    \includegraphics[scale=0.5]{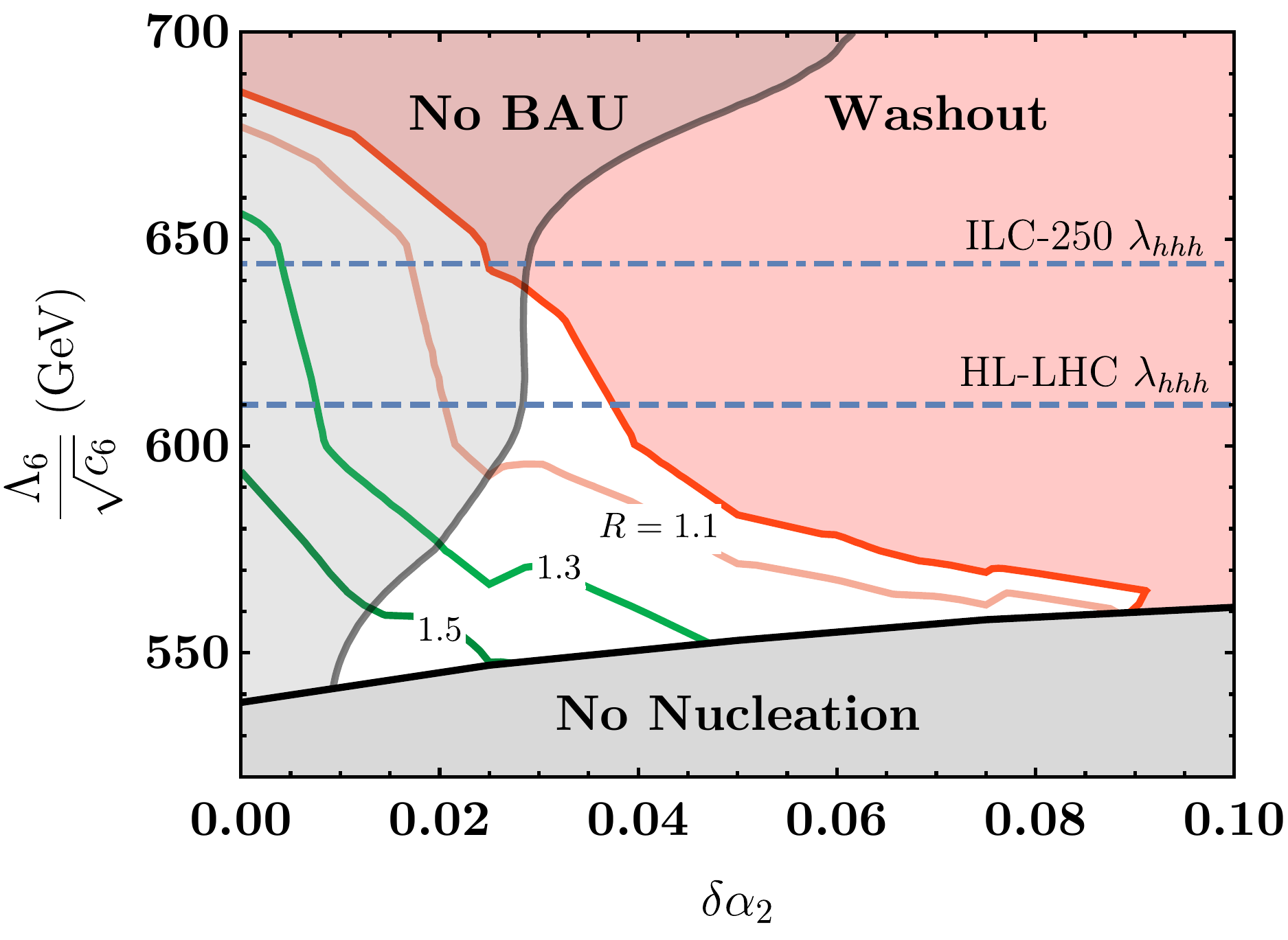} 
    \caption{Viable parameter space as a function of $\deltaW$ and $\Hcut$, for $\delta\alpha_3 = 0$. The grey region at lower values of $\Hcut$ is ruled out due to bubbles of true vacuum not being able to nucleate. The transparent grey region on the left shows where an insufficient BAU is produced given $\Lambda_{\rm CPV}/\sqrt{\delta_{\rm CPV}}=7.3$~TeV. The red region is where sufficient BAU is produced outside of the true vacuum, but the phase transition is too weakly first order and hence the EW sphalerons wash out the BAU in the broken phase. The red, pink, light green and dark green contours correspond to values of the ratio $R$, defined in Eq. \Cref{xi.EQ}, of $1$, $1.1$, $1.3$ and $1.5$ respectively.  Also shown are contours corresponding to the $68\%$ CL bounds from HL-LHC (dashed) and ILC-250 (dot-dashed), as found in \cite{DiVita:2017vrr}.}
    \label{fig:EWBG}
\end{figure}

\begin{figure}[t]
    \centering
    \includegraphics[scale=0.5]{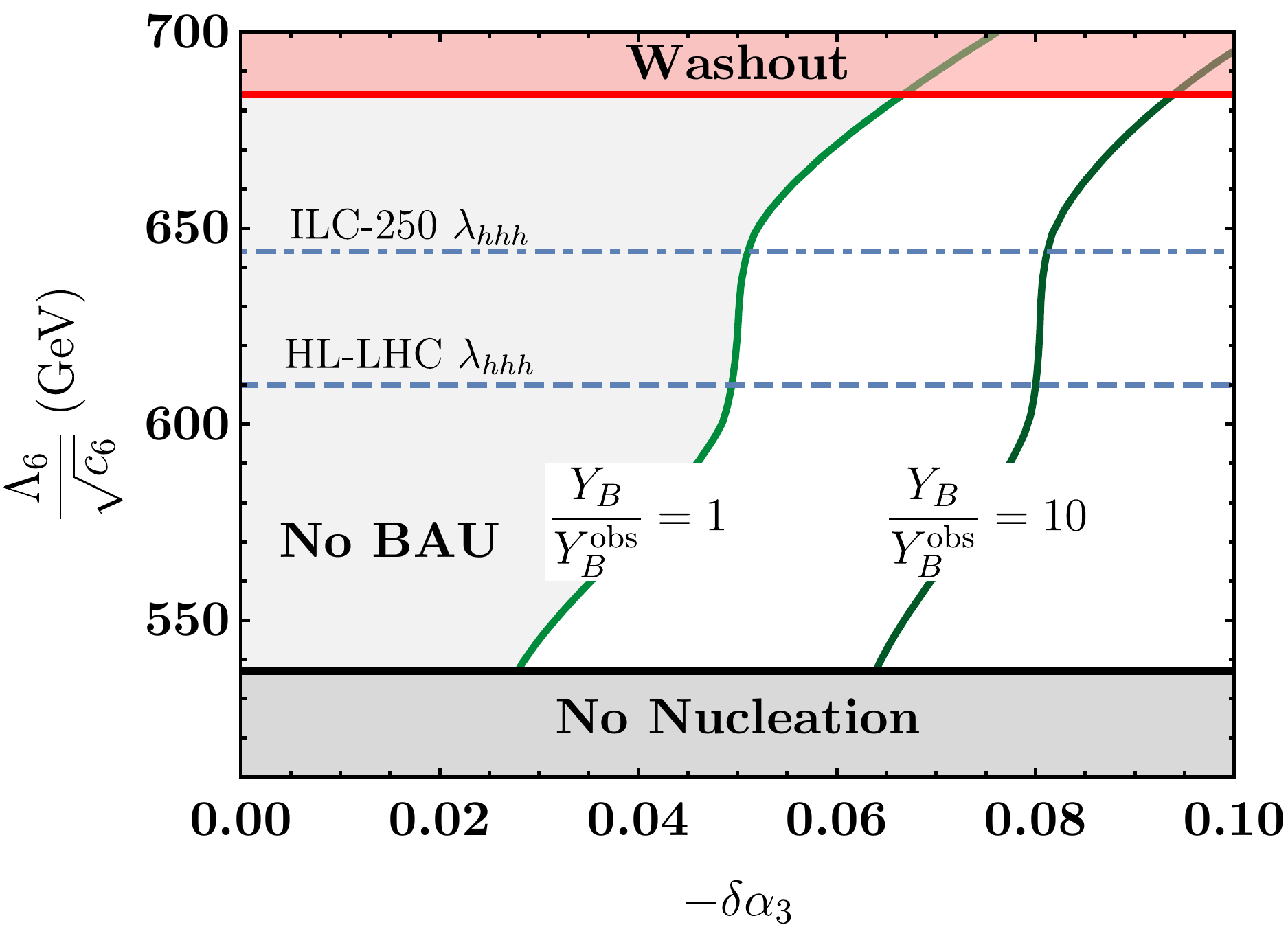} 
    \caption{Viable parameter space as a function of $\delta\alpha_3$ and $\Hcut$ for $\deltaW=0$. The grey region at lower values of $\Hcut$ is ruled out due to bubbles of true vacuum not being able to nucleate. The transparent grey region on the left shows where an insufficient BAU is produced. The red region is where an initially sufficient BAU is produced, but the phase transition is too weakly first order, so that the EW sphalerons wash out the BAU in the broken phase, assuming a washout factor of $F_W = 10\%$, defined in Eq. \Cref{fw.EQ}. The light and dark green contours correspond to obtaining the correct BAU, and a factor of ten more than required, respectively. Also shown are contours corresponding to the $68\%$ CL bounds from HL-LHC (dashed) and ILC-250 (dot-dashed), as found in \cite{DiVita:2017vrr}.}
    \label{fig:EWBGalpha3}
\end{figure}

Electroweak baryogenesis with \CP violation entering through the top Yukawa coupling  and an $|H|^6$ operator to induce a first order phase transition is ruled out by the latest electron EDM constraint~\cite{Andreev:2018ayy}, $\Lambda_{\rm CPV}/\delta_{\rm CPV}>7.3$~TeV. However, if the efficiency of producing a baryon asymmetry can be otherwise enhanced, this scenario can be revived. In order to make this point more clear, we set
\begin{align}
\frac{\Lambda_{\rm CPV}}{\sqrt{\delta_{\rm CPV}}}=7.3~{\rm TeV}.
\end{align} 
In \Cref{CPV.FIG} we show how the required scale of \CP violation changes by varying the weak and strong coupling constants.

Altering the gauge couplings in the early universe can dramatically alter the efficiency for producing a baryon asymmetry by changing the weak and strong sphaleron rates. 
In a first-order phase transition, bubbles of true vacuum (where the EW symmetry is broken) nucleate and grow. Outside of these bubbles, the EW symmetry remains unbroken. \CP\hspace{-0.1cm}-violating interactions between the quarks and the bubble wall catalyze a chiral asymmetry in front of the bubble. Due to its large Yukawa coupling, the most important particle involved in these interactions is the top quark. This chiral asymmetry biases the EW sphalerons to create a baryon asymmetry in front of the bubble wall, some of which is swept up into the broken phase by the expanding bubble. Since the EW sphalerons are responsible for generating the baryon asymmetry, enhancing them in the unbroken phase can lead to a greater final baryon asymmetry. A detailed discussion of the effect of this enhancement persisting in the broken phase can be found in  \Cref{washout.SEC}. In order for this process to be efficient, the chiral asymmetry produced in front of the bubble wall should not be washed out by strong sphalerons. Thus, while increasing the EW sphaleron rate in the unbroken phase would increase the baryon asymmetry, likewise a suppression of the strong sphaleron rate would lead to less washout of the chiral asymmetry, and therefore a greater baryon asymmetry. 

In order to get the most accurate results, sphaleron/instanton rates should be calculated using non-perturbative techniques such as on a lattice. However analytical approximations exist for both the weak~\cite{Bodeker:1999gx} and the strong sphaleron rates \cite{Moore:2010jd} in thermal equilibrium and describe the underlying phenomena to a good degree,
\begin{align}
\Gamma _{\rm WS}\simeq120\, \alpha _2^5\, T\quad {\rm and}\quad \Gamma _{\rm SS} \simeq132\, \alpha _3^5\, T~. \label{eq:Gsph}
\end{align}
The weak sphaleron rate is given here in the symmetric phase, and is suppressed by an exponential factor $\sim \exp(-M_W/T)$ in the broken phase. \Cref{eq:Gsph} shows the sensitivity of the strong and weak sphaleron rates to the gauge couplings. We emphasize that the main effect of modifying the gauge couplings is to change these sphaleron rates. While we apply this variation of gauge couplings to a particular model in this paper, the inclusion of the relevant dimension-5 operators of  \Cref{eq:model} in any other model of EW baryogenesis can also be investigated. We leave such studies to future work.

In computing the final baryon asymmetry, we start by finding the profile of the true vacuum bubble. This step is required because we invoke a non-renormalizable operator as the source of \CP violation, so that the baryon asymmetry is sensitive to the bubble wall width \cite{Balazs:2016yvi}. We obtain the bubble wall profile by solving the classical equations of motion of the Higgs field across the spatial boundary between regions of true and false vacuum. This solution is a smooth function, and can be fit to a $\tanh(x)$ function for ease in the remainder of the calculation. We then make use of the vev-insertion approach outlined in \cite{Lee:2004we} to calculate the chiral asymmetry produced from \CP\hspace{-0.1cm}-violating interactions with the bubble wall.\footnote{The accuracy of the vev-insertion approach is yet to be thoroughly tested and it remains unclear whether it results in an under- or over-estimate of the baryon asymmetry. However, a more rigorous treatment of a toy model did confirm the existence of a resonance-like feature \cite{Cirigliano:2011di}. Therefore, numerical values of the baryon asymmetry we find should be understood as being indicative of the dependence of the BAU on the parameters of the theory, as opposed to a precise calculation of the final baryon abundance. } We then calculate the resultant baryon asymmetry produced by the $SU(2)_L$ sphalerons as an integral over the left-handed fermion density $n_L$  \cite{Carena:2000id,Cline:2000nw}
\begin{equation}
    Y_B = \frac{3 \Gamma _{\text{WS}}(\alpha_2)}{2  s D_Q(\alpha_i) \kappa _+(\alpha_i)}  \int _{-\infty} ^0 \text{d}y\ n_L (\alpha_i, y)\ e^{-\kappa _-(\alpha_i) y}\ .
    \label{YB.EQ}
\end{equation}
The quantity $\kappa_\pm$ is defined as
\begin{equation}
\kappa _\pm  = \frac{v_w \pm \sqrt{v_w^2 +15 D_Q(\alpha_i) \Gamma _{\text{WS}}(\alpha_2) }}{2D_Q(\alpha_i)} \ , 
\end{equation}
where $v_w$ is the wall velocity, $D_Q(\alpha_i)$ is the diffusion coefficient which depends on $\alpha_i$ (see \Cref{eq:DQ}), $\Gamma _{\text{WS}}$ is the $SU(2)_L$ sphaleron rate, and $s$ is the entropy density. A detailed discussion of how we compute $Y_B$ is given in \Cref{sec:appendix}.

From  \Cref{YB.EQ}, we can observe that increasing the weak sphaleron rate in the unbroken phase will increase the final baryon yield.  The growth of $Y_B$ with $\alpha_2$ is quite dramatic since the weak sphaleron rate grow as $\alpha_2^5$ . However, the enhancement from modified sphaleron rates will not diverge for the following two reasons. First, if we assume an exponential profile for $n_L(z)$ in Eq. \Cref{YB.EQ}, a linear term in $\Gamma _{\text{WS}}$ exists in both the numerator and denominator, so that for large enough $\Gamma _{\text{WS}}$ the yield $Y_B$ asymptotes. In addition, while initially increasing $\alpha _2$ raises the masses of the $SU(2)_L$ doublets $m_L$ such that \emph{CP}-violating interactions with the bubble wall approach a resonance, eventually $\alpha _2$ is so large that $m_L>m_R$, which leads to non-resonant interactions (further details given in the appendix around \Cref{eq:thermmass}. Finally, $\alpha _2$ modifies the bubble wall profile which thereby changes the profile of the \emph{CP}-violating source. More details of how modifications of $\alpha_2$ enter every step of the calculation are provided in \Cref{sec:appendix}.

The dependence of the baryon asymmetry on $\alpha _3$ mostly arises from the suppression of the strong sphaleron rate.  The strong sphaleron rate relaxes the chiral asymmetry. Therefore, if the strong sphaleron rate is decreased, then $n_L$, the chiral asymmetry, increases. Again this effect is quite dramatic due to the $\alpha _3^5$ scaling of the strong sphaleron rate. In addition the reduction in the strong coupling increases the diffusion coefficient which tends to moderately enhance the BAU. Finally this growth of the BAU is resisted slightly by the fact that the \emph{CP}-violating sources are largest for $m_L=m_R=2T$ and the thermal masses are decreasing with smaller $\alpha _3$. We again relegate further details to \Cref{sec:appendix}.

We emphasize that the result of \Cref{YB.EQ} is the baryon asymmetry produced in the unbroken phase and in front of the bubble wall, without accounting for dynamics in the broken phase of EW symmetry. We discuss in the next section how this asymmetry can be washed out if weak sphalerons are still active in the broken phase, inside the bubble. 

\subsection{Avoiding Washout}
\label{washout.SEC}

\begin{figure}[t]
\centering
\includegraphics[scale = 0.45]{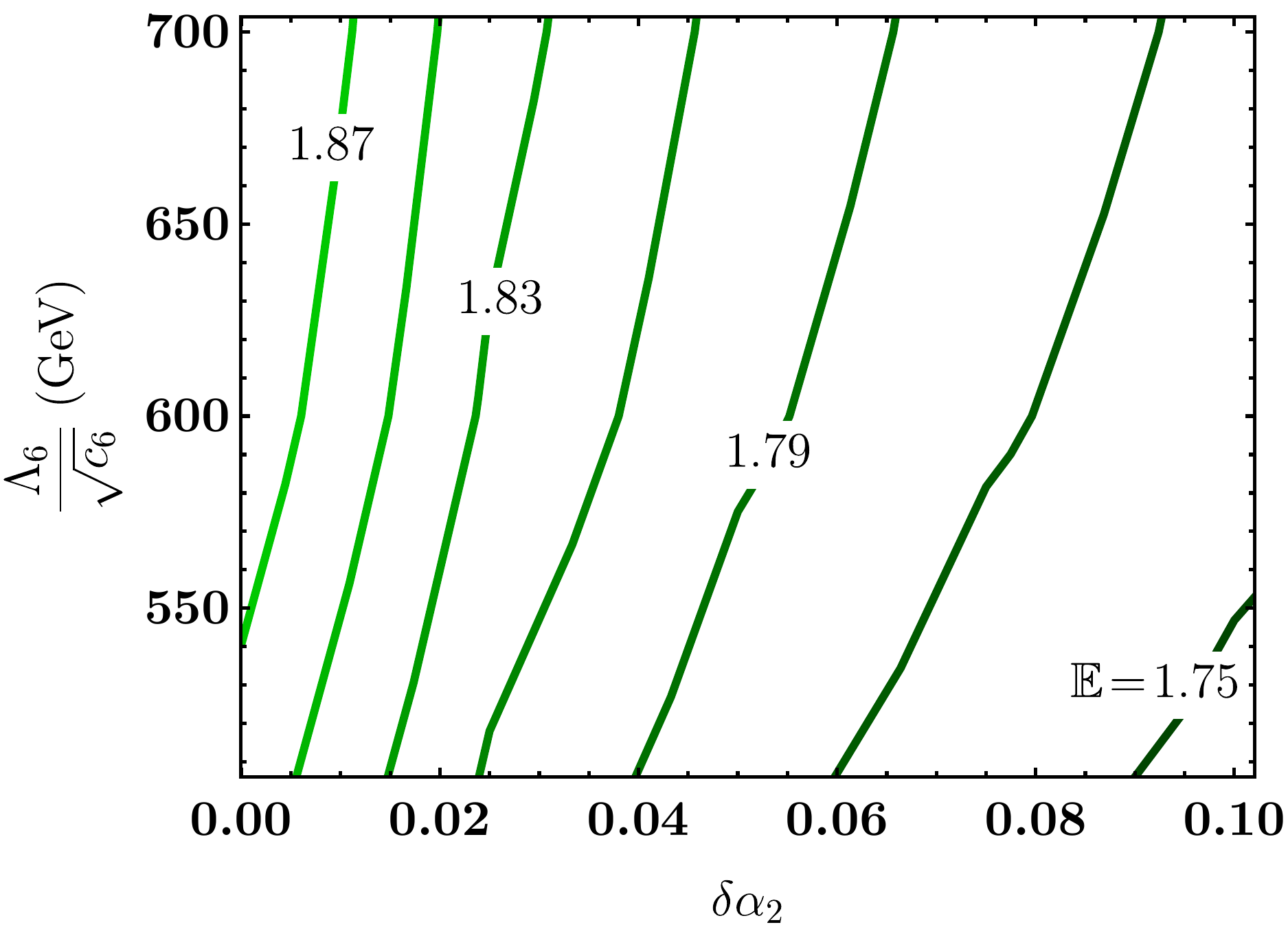}
\caption{Contours of the sphaleron energy in units of $4 \pi v/ g_2$ as a function of both $\deltaW$ and $\Hcut$, as defined in \Cref{eq:Esph}.}
\label{EsphPlot.FIG}
\end{figure}

The success of the mechanism presented here depends strongly on the enhanced sphaleron rate in the unbroken phase of EW symmetry and across the bubble walls. It therefore also depends strongly on ensuring that the enhanced sphaleron rate, if it persists in the broken phase inside the bubble, does not wash out the generated baryon asymmetry. In order for this washout to be prevented, it is clear that the phase transition should be strongly first order. This would ensure that the change in the sphaleron rate between the unbroken and broken phase is drastic, so that the enhanced sphaleron rate is nevertheless small inside the bubble of true EW vacuum.

Inside the bubble of true vacuum, baryon number density $n_B$ is depleted according to
\beq
\partial_t n_B = -\frac{13 N_F}{2}\frac{\Gamma_{\rm sph}(T)}{V T^3} n_B \propto -e^{-E_{\rm sph} /T} n_B \ ,
\eeq
where $E_{\rm sph}$ is the sphaleron energy. Given a phase transition duration $t_{\rm PT}$, we may define a washout factor as given by 
\beq
F_W = \frac{n_B(t_{\rm PT})}{n_B(0)} \ ,
\label{fw.EQ}
\eeq
which will provide a bound on the degree to which washout is tolerated in a model of baryogenesis. This bound can be incorporated into an approximate baryon number preservation condition (BNPC) as a function of $v_c/T_c$~\cite{Patel:2011th}, which is 
\beq
\frac{4 \pi \mathbb{E}(\lambda_H, g_2, c_6)}{g_2}\frac{v_c}{T_c} - 6 \log \frac{v_c}{T_c} > -\log (-\log F_W) - \log \frac{t_{\rm PT}}{t_H} + \log  \chi_N+ \log \kappa \ ,
\label{BNPC.EQ}
\eeq
where 
\begin{align}
\chi_N &= \left[ \left( \frac{13 N_F}{2}\right) \mathcal{N}_{\rm tr} (\mathcal{N} \mathcal{V})_{\rm rot} \left( \frac{\omega_- t_H}{2\pi}\right)\right] \ , 
\end{align}
$t_H = 1/H(T)$ is the Hubble time, $\kappa$ is the fluctuation determinant ratio that goes into calculating the EW sphaleron rate, $\omega_-$ is the frequency of the unstable mode of the sphaleron, $N_F$ is the number of fermion families, and $\mathcal{N}_{\rm tr} (\mathcal{N}\mathcal{V})_{\rm rot} \sim 7000$ are translational and rotational model factors which also enter the sphaleron rate and were computed in \cite{Carson:1989rf}. 

The fluctuation determinant ratio $\kappa$ is a sensitive function of $\lambda_H / g_2^2$, and was computed in \cite{Carson:1990jm} for four values of $\lambda_H / g_2^2 = 0.1,~0.3,~1,~10$, from which an extrapolation in the range $5\times10^{-2} \lesssim \lambda_H / g_2^2 \leq 10$ was presented. In that analysis, it was found that for $\lambda_H / g_2^2 \sim 0.3$ as in the SM, $\log \kappa \sim -2.2$, while for $\lambda_H / g_2^2 \sim 0.1$, $\log \kappa \sim -7.9$. Since our model increases $g_2$, it pushes us to values of $\lambda_H / g_2^2 <0.3$ and we enter a regime where $\log\kappa$ is quite sensitive to the precise choice of $\delta\alpha_2$. The largest deviation from the SM we will consider is $\delta\alpha_2 \sim 0.1$, which corresponds to $\lambda_H / g_2^2 \sim 0.08$. This is at the lower limit of the interpolating curve shown in \cite{Carson:1990jm}, which gives $\log \kappa \sim -11$. Since an explicit numerical calculation of the fluctuation determinant ratio as a function of $\delta\alpha_2$ is beyond the scope of our study, and the calculation of \cite{Carson:1990jm} was only performed for four specific values, our choice of $\log \kappa$ has an inherent uncertainty which we are currently unable to quantify. However, the perturbative calculation of $\log\kappa$ gives substantially larger values than the numerical results of \cite{Carson:1990jm}, so we can hope that the true value of $\log\kappa (\delta\alpha_2)$ is not smaller than that which we use, meaning that our estimate might be conservative.

In our analysis of the washout avoidance condition, we will take 
\beq
-\log F_W \sim \text{min}[ 0.1,~ \log Y_B/Y_B^{\rm obs} ] \ ,
\eeq 
which corresponds to a washout factor of either $10\%$ or the amount by which our mechanism overproduces a baryon asymmetry. We use this washout factor because we find that for large $\delta\alpha_2$ we can obtain a substantially larger BAU than is required, and therefore tolerate a significantly larger washout factor. 

The function $\mathbb{E}(\lambda_H, g_2, c_6)$ appearing in \Cref{BNPC.EQ} is the sphaleron energy in units of $4 \pi v / g_2$. This function can be obtained only by computing the sphaleron solution of the effective field theory by starting with the usual \textit{ansatz} of \cite{Klinkhamer:1984di} for the gauge and scalar fields
\beq
W_i^a \sigma^a  = -\frac{2i}{g_2}f(\xi) \partial_i U U^{-1} ,~~ H = \frac{v}{\sqrt{2}}h(\xi) U\begin{pmatrix} 0 \\ 1 \end{pmatrix} \ ,
\eeq
where $U$ is an element of $SU(2)$, and $\xi = g_2 v r$ is a dimensionless radial coordinate. In the presence of the $\frac{c_6} {\Lambda_6^{2}} |H|^6$ operator, the usual coupled non-linear differential equations of motion for $f(\xi)$ and $h(\xi)$ are modified and can be written as \cite{Grojean:2004xa}
\begin{align}
\nonumber \xi^2 \frac{d^2 f}{d\xi^2} &= 2 f(1-f) (1-2f) - \frac{\xi^2}{4}h^2(1-f) \ , \\
\frac{d}{d\xi}\left(\xi^2 \frac{d h}{ d\xi} \right) & = 2h(1-f)^2 + \frac{\lambda_H}{g_2^2} \xi^2(h^2-1) h + \frac{3}{4}\frac{v^2 c_6 }{g_2^2 \Lambda_6^2}\xi^2 h(h^2-1)^2 \ .
\end{align}
We solve these differential equations numerically using the Newton-Kantorovich Method (NKM) as described in great detail in \cite{Gan:2017mcv}. 

The sphaleron energy at $T=0$ is given by $E_{\text{sph}} = \int d^3 x T^{00}$, where $T^{\mu\nu}$ is the stress-energy tensor, and may be written as
\begin{align}
\nonumber E_{\text{sph}} &= \frac{4 \pi v \mathbb{E}(\lambda_H, g_2, c_6)}{g_2} \\
\nonumber &= \frac{4 \pi v }{g_2} \int_0^{\infty} d\xi \Bigg( 4 \left( \frac{df}{d\xi}\right)^2 + \frac{8}{\xi^2}f^2(1-f)^2 + \frac{1}{2}\xi^2\left( \frac{dh}{d\xi}\right)^2 \\
&~~~~~~~~~~~~~~~~~~~~~~~~+ h^2(1-f)^2 + \frac{\lambda_H}{4 g_2^2} \xi^2(h^2-1)^2 + \frac{v^2 c_6 }{8 g_2^2 \Lambda_6^2} \xi^2(h^2-1)^3 \Bigg) \ . \label{eq:Esph}
\end{align}
In practice, we compute this integral numerically from the solutions $h(\xi)$ and $f(\xi)$ obtained using the NKM. We use a symmetric numerical integration procedure, where 
\beq
\frac{d X}{d \xi} = \frac{X_{i+1} - X_{i-1}}{2 \Delta \xi} \ ,
\eeq
where $X_i= f_i,~h_i$ corresponds to the value of the function at step $i$, and $\Delta \xi$ is the separation between steps. The result for $\mathbb{E}(\lambda_H, g_2, c_6)$ is shown in  \Cref{EsphPlot.FIG}. The behaviour is as expected, in that at large $\delta \alpha_2$, or correspondingly, small $\lambda_H / g_2^2$, the normalised sphaleron energy asymptotes towards a fixed value $\mathbb{E} \sim 1.6$ as $g_2 \to \infty$. For larger suppression scales than those shown in  \Cref{EsphPlot.FIG}, the sphaleron energy increases (because we have chosen $c_6 >0$) until we recover the SM Higgs sector in the limit $\Lambda_6 \to \infty$ \cite{Grojean:2004xa, Gan:2017mcv}.

The sphaleron energy is used in the calculation of the BNPC of \Cref{BNPC.EQ}. This enables us to compute the required value of $v_c/T_c$ that ensures that sphaleron rates are sufficiently suppressed inside the bubbles of true vacuum, so that our produced baryon asymmetry is not washed out. This requirement must then be contrasted with what is actually obtained in our model. We define the ratio
\beq
R = \frac{v_c/T_c }{v_c/T_c}\begin{matrix}|_{\rm obt.} \\ |_{\rm req.} \end{matrix} \ ,
\label{xi.EQ}
\eeq
to parameterize the extent to which the baryon asymmetry produced by the enhanced out-of-equilibrium sphalerons is not washed out by the sphalerons remaining enhanced in the broken phase. The values of $v_c/T_c$ required range between $1.4 \lesssim v_c/T_c \lesssim 2.8$, with the larger values being required at large $\deltaW$. Meanwhile the values of $v_c/T_c$ obtained are typically no larger than $v_c/T_c \sim 2$, and the largest values are obtained for low $\Hcut$ and low $\deltaW$, as can be seen in \Cref{fig:vcTc}.

\subsection{Discussion of Results}
Due to the large number of moving parts in the calculation of the final baryon asymmetry obtained when varying gauge couplings, we discuss here briefly the main results, which are shown in \Cref{CPV.FIG,fig:EWBG,fig:EWBGalpha3}.

In  \Cref{CPV.FIG}, a particular choice of $\Hcut = 575$ GeV was made to show how the scale of new \CP violation that yields the correct baryon asymmetry varies as a function of $\deltaW$. As mentioned, new \CP violation in the top quark sector is ruled out by non-observation of the electron EDM if $\deltaW = \delta\alpha_3 = 0$. This can be seen by the light blue contour only appearing for values of $\deltaW \gtrsim 0.02$. The location of these contours would shift downwards for higher values of $\Hcut$. All of this parameter space could potentially be probed in future electron EDM searches \cite{2018PhRvL.120l3201L,Vutha:2017pej,Kozyryev:2017cwq}.

In \Cref{fig:EWBG}, we show the viable parameter space when we vary only $\deltaW$ and $\Hcut$, taking $\delta\alpha_3=0$. Here, three effects compete to reduce the viable regions. On the one hand, if $\Hcut$ is too low, the tunneling rate is too low, and bubbles of true vacuum cannot nucleate, as shown in the lower grey shaded region. On the other, the observed baryon asymmetry cannot be reproduced if $\deltaW$ is too small, due to the constraint from the electron EDM on the scale of new sources of \CP violation. Finally, while a large baryon asymmetry can be produced by the enhanced sphalerons in a wide region of the parameter space (above and to the right of the grey shaded regions), it is also washed out by the other side of the double-edged sword that is a higher sphaleron rate. Indeed, as one increases $\deltaW$, the baryon asymmetry grows rapidly, and is often overproduced. However, if the sphaleron rate remains enhanced in the broken phase, for too large $\deltaW$ or $\Hcut$, the strength of the FOPT is insufficient to prevent washout of this initial overproduction. Thus, the viable parameter space is restricted to certain values of $v_c/T_c$ which lie below the red contour of $R = 1$ (See \Cref{xi.EQ}).

In \Cref{fig:EWBGalpha3}, we show the parameter space when varying only $\delta\alpha_3$ and $\Hcut$, given $\deltaW=0$. The region in which bubbles of true vacuum do not nucleate is determined by our choice of $\deltaW$, such that for positive non-zero values of $\deltaW$, this exclusion region would shift upwards. The region in which the initial BAU cannot be achieved would shift to the left if $\deltaW > 0$ were allowed. Meanwhile, the region ruled out by sphaleron washout would be lower for fixed washout factor $F_W \sim 10\%$ if $\deltaW$ were varied. This choice corresponds to a conservative estimate, since this constraint can in principle be relaxed if we were to take advantage of the initial overproduction of the baryon asymmetry.

We do not present a combined analysis of simultaneous variations of $\alpha_2$ and $\alpha_3$. As will be discussed in the next section, the UV completions envisaged to modify the gauge couplings in the early universe tend to not apply simultaneously to both $\alpha_2$ and $\alpha_3$. 

Interestingly, the vast majority of the open parameter space can be probed by constraining the triple-Higgs coupling $\lambda_{hhh}$ at High Luminosity LHC (HL-LHC), and all of it (if only $\alpha_2$ is varied) could be probed at the ILC running at $\sqrt{s} = 250$ GeV \cite{DiVita:2017vrr}. Shown in \Cref{fig:EWBG,fig:EWBGalpha3} are 68\% CL limits on $\Hcut$ as obtained from a global fit, which for ILC assume a combination with HL-LHC. The 95\% CL bounds for HL-LHC do not appear in the figure shown. The proposals to run ILC at $\sqrt{s}=250,~350, 500$ GeV, or the FCC-ee at $\sqrt{s} = 240,~350$ combination could place 95\% CL bounds on the entire parameter space shown in \Cref{fig:EWBG,fig:EWBGalpha3}.

\section{Models for Modifying the Gauge Couplings}
\label{sec:varyingcouplings}
In this section we discuss two different mechanisms for generating the dimension-5 operators in the effective Lagrangian in \Cref{eq:model} which control the size of the various gauge couplings at temperatures near that of the EW phase transition. We discuss how each mechanism can only serve to modify one of either the weak or strong gauge coupling at a time, with experimental constraints preventing the simultaneous modification of the other. The first mechanism, which involves an ultra-light scalar with a non-zero energy density which scan values of the gauge couplings, additionally requires modifying the hypercharge gauge coupling so as to evade constraints during Big Bang Nucleosynthesis and from measurements of stars at $\mathcal{O}(1)$ redshifts. The second mechanism requires the existence of an additional scalar field near the EW phase transition scale which obtains a vev. In order for this scalar to couple to gauge bosons sufficiently, a large number of new states must also exist near that scale, leading to possible experimental signatures.

\subsection{Scanning Coupling Constants}
\label{sec:scanningcouplings}

We first discuss the first mechanism, whereby an ultra-light scalar field, $\p$, scans values of the gauge couplings as it evolves in time. Such a scalar field may be treated as a coherently oscillating classical field whose energy density behaves like non-relativistic matter as long as its potential is dominated by the $\p^2$ term \cite{Turner:1983he}. This ultra-light scalar could be a light modulus or dilaton-like field, which we will henceforth treat as making up some fraction of the Dark Matter (DM) energy density $\rho_{\text{DM}}$, denoted $f_{\text{DM}}$. We therefore write the field as
\begin{equation}
    \p \simeq \frac{\sqrt{2 f_{\text{DM}} \rho_{\text{DM}}}}{m_\p}\cos\left(m_\p  t \right) \ ,
    \label{phi.EQ}
\end{equation}
where $m_\p$ is the mass of the $\p$ field.

If this field has couplings to the field strengths of the SM gauge groups, we may write these as
\begin{align}
   \nonumber \Lag \supset -&\frac{1}{4g_Y^2}\left( 1-\frac{c_{g_Y} \p}{\MPl}\right)  B^{\mu \nu}{B}_{\mu \nu}-\frac{1}{4g_2^2}\left( 1-\frac{c_{g_2} \p}{\MPl}\right)  W^{a, \mu \nu}{W}^a_{\mu \nu}\\
    &-\frac{1}{4g_3^2}\left( 1+\frac{2c_{g}\, g_3\, \beta_3\, \p}{\MPl}\right)  G^{A, \mu \nu}{G}^A_{\mu \nu} \ ,
    \label{ScalarCouplings.EQ}
\end{align}
where $B_{\mu\nu},~{W}^a_{\mu \nu},~{G}^A_{\mu \nu}$ are the $U(1)_Y,~SU(2)_L,~SU(3)_c$ field strength tensors respectively and we have set $\Lambda_{Y,2,3}$ in \Cref{eq:model} to a slightly modified value of the Planck mass, $\MPl = (4\pi G_N)^{-1/2} = 3.4\times 10^{18}$~GeV, to match the literature on dilaton couplings.\footnote{Note that with our definition, $\MPl = \sqrt{2} m_\text{P}$, where $m_\text{P}$ is the usual reduced Planck mass.} The normalizations of the scalar couplings $c_{g_Y}$, $c_{g_2}$, $c_{g}$ are also chosen to match the treatment of dilaton couplings  \cite{Kaplan:2000hh,Damour:2010rp,Damour:2010rm}. This choice of normalization requires a coupling of the scalar to fermion masses
\begin{equation}
    \Lag \supset - \frac{c_g \p}{\MPl}\sum_{f} \gamma_{m_f}  m_f \bar{\psi}_f \psi_f \ ,
\end{equation}
such that the coupling $c_g$ appears in a RG-invariant manner.\footnote{The RG-invariance holds only up to electromagnetic corrections which are $\alpha_{EM}$ suppressed.} This formulation of the coupling $c_g$ is such that it is equivalent to coupling $\p$ to the anomalous part of the gluon stress-energy tensor, which in turn ensures that $c_g$ is a measure of the coupling of $\p$ to the gluonic energy component of the mass of a hadron.

These couplings of the ultra-light scalar to the gauge kinetic terms amount to a change in the effective gauge coupling, now given by
\begin{equation}
    g_{i,\text{eff}} = \frac{g_i}{(1-\frac{c_{g_i}\, \p}{\MPl})^{1/2}} \ ,
    \label{EffCoupling.EQ}
\end{equation}
for the $U(1)_Y$ and $SU(2)_L$ couplings. The $SU(3)_c$ coupling modification is not written out because as we will discuss later, constraints on $c_g$ are such that no substantial modification of $\alpha_3$ will be possible by this mechanism. Indeed, the constraint is such that only for $m_\p \lesssim 10^{-32}$ eV can one achieve $\delta \alpha_3 \gtrsim 0.03$ as required for successful baryogenesis in our model. This value of $m_\p$ is barely compatible with the picture of a coherently oscillating scalar field, which requires $m_\p \gtrsim 3H_0 \sim 10^{-33}$ eV, where $H_0$ is the value of the Hubble constant today \cite{Frieman:1995pm}. Since this is at the boundary of compatibility with constraints, we will only consider modification of $\alpha_Y$ and $\alpha_2$ in this discussion.

If the scalar $\p$ behaves as non-relativistic matter, its energy density will evolve as $T^3$ from $T_{\text{eq}} \sim  \left(m_\p M_{\text{Pl}} \right)^{1/2}$ to the temperature now, $T_{\text{0}}$, so that the field value of $\p$ was $(T/T_{0})^{3/2}$ greater at a temperature $T<T_{\text{eq}}$ than it is now. Therefore, $g_{i,\text{eff}}$ will vary as a function of the temperature between now and $T_{\text{eq}}$. At temperatures above $T_{\text{eq}}$, the value of the scalar $\p$ behaves as dark energy would, and therefore does not change. In this section, we only consider scalars with masses below $m_\p \ll 10^{-6}$ eV, i.e. $T_{\text{eq}} \ll T_{\text{EWPT}}$, so that during the electroweak phase transition the scalar $\p$ is not undergoing oscillations.

\subsubsection{Constraints on Scalar Couplings}
The couplings of ultra-light scalars have been strongly constrained by searches for long-range Equivalence Principle (EP) preserving ``fifth forces" and for EP-violating forces. They have also been strongly constrained by searches for variations in the fundamental constants.

At low scalar masses, the strongest constraint on the coupling $c_g$ comes from a search for deviations in the gravitational lensing of the sun, performed by the Cassini spacecraft \cite{Bertotti:2003rm}. This search places a constraint on the Eddington parameter $\gamma$:
\begin{equation}
    1-\gamma \simeq 2 c_g^2 \leq \left(2.1 \pm 2.3\right)\times 10^{-5} \ ,
\end{equation}
such that $|c_g| \lesssim 6\times 10^{-3}$ at 2$\sigma$. Hence any modifications of the $SU(3)_c$ coupling will be minimal, as claimed above.

The searches for long-range forces are conducted at energies well below the EW symmetry breaking scale, and so they typically refer to the coupling to the photon kinetic term as opposed to separating into the $U(1)_Y$ and $SU(2)_L$ kinetic terms. Thus they constrain a coupling $c_e$, which appears in the Lagrangian as
\begin{equation}
   \Lag \supset -\frac{1}{4e^2}\left( 1-\frac{c_{e}\,  \p}{\MPl}\right)  A^{\mu \nu}{A}_{\mu \nu} \ .
\end{equation}
Using the relationship between electromagnetic (EM) coupling $e$ and the $U(1)_Y$ and $SU(2)_L$ gauge couplings,
\begin{equation}
    e = \frac{g_Y g_2}{(g_Y^2 + g_2^2)^{1/2}} \ ,
\end{equation}
we can express the coupling of the scalar to the EM kinetic term in terms of the couplings $c_{g_2}$ and $c_{g_Y}$ as 
\begin{equation}
    c_e =  \alpha_Y c_{g_2} + \alpha_2\, c_{g_Y} \ .
    \label{deCouplingRel.EQ}
\end{equation}
In turn, we may now discuss the constraints that have been set on variations of the EM fine structure constant $\alpha_{\text{EM}}$ in terms of constraints on the scalar couplings to the $U(1)_Y$ and $SU(2)_L$ kinetic terms. The variation of the EM coupling as a function of the scalar field value $\p$ is
\begin{equation}
 \left( \frac{\Delta \alpha_{\text{EM}}}{\alpha_{\text{EM}}}\right) = \frac{c_e\, \p}{\MPl-c_e\, \p} \ ,
\end{equation}
which will enable us to constrain $c_e$, and in turn $\dgy$ and $\dgw$. 

There exist strong constraints on the variation of the EM coupling from astrophysical data, taken at redshifts between $1 < z < 4$. A meta-analysis of recent measurements yields a weighted mean \cite{Martins:2017yxk} 
\begin{equation}
    \left( \frac{\Delta \alpha_{\text{EM}}}{\alpha_{\text{EM}}}\right)_{\text{astro, '17}} =(-0.64 \pm 0.65)\times10^{-6} \ ,
\end{equation}
which is consistent with zero at $1\sigma$, while a previous analysis \cite{Webb:2010hc} found 
\begin{equation}
    \left( \frac{\Delta \alpha_{\text{EM}}}{\alpha_{\text{EM}}}\right)_{\text{astro, '10}} =(-2.16 \pm 0.86)\times10^{-6} \ ,
\end{equation}
which indicated a possible variation of the EM coupling at high redshift, but remained consistent with zero at a little over $2\sigma$. The first average above for $\Delta \alpha_{\text{EM}}/\alpha_{\text{EM}}$ corresponds to
\begin{equation}
c_e \sim -8\pm 8 \times 10^{25}\, \left(\frac{m_\p}{\text{eV}}\right) f_{\text{DM}}^{-1/2} \ ,
\end{equation}
assuming a central observation redshift of $z=2.5$ and a gravitationally-suppressed $\p$ interaction.

Additionally, there is a strong constraint on the variation of the EM coupling from the Oklo natural fission reactor \cite{Damour:1996zw}, which constrains
\begin{equation}
-0.9\times10^{-7}< \left( \frac{\Delta \alpha_{\text{EM}}}{\alpha_{\text{EM}}}\right)_{\text{Oklo}} < 1.2\times10^{-7} ,
\end{equation}
which also constrains the annual variation of the EM coupling to be less than one part in $\sim10^{17}$. This constraint corresponds to a constraint
\begin{equation}
-6 \times 10^{25}  \lesssim c_e \ \left(\frac{m_\p}{\text{eV}}\right)^{-1} f_{\text{DM}}^{1/2} \lesssim 8 \times 10^{25}  \ ,
\end{equation}
for a gravitationally-suppressed $\p$ interaction as above.

There are also constraints on the variation of $\alpha_{\text{EM}}$ from Big Bang Nucleosynthesis (BBN), which can be of similar strength if the variation is coupled to variations of other fundamental parameters \cite{Coc:2006sx}. However, since these coupled variations are not considered in this paper, we will not discuss them further.\footnote{One might be concerned that BBN should constrain variations in $\alpha_2$ due to the sensitivity of the final Helium abundance to the neutron lifetime, which is a weak process. The neutron decay rate can be written so that $\Gamma_n \propto (g_2/M_W)^4$, so that it is sensitive to the Higgs VEV during BBN, but not $g_2$ directly, at tree level. }

\begin{figure}[t]
    \centering
    \includegraphics[scale=0.5]{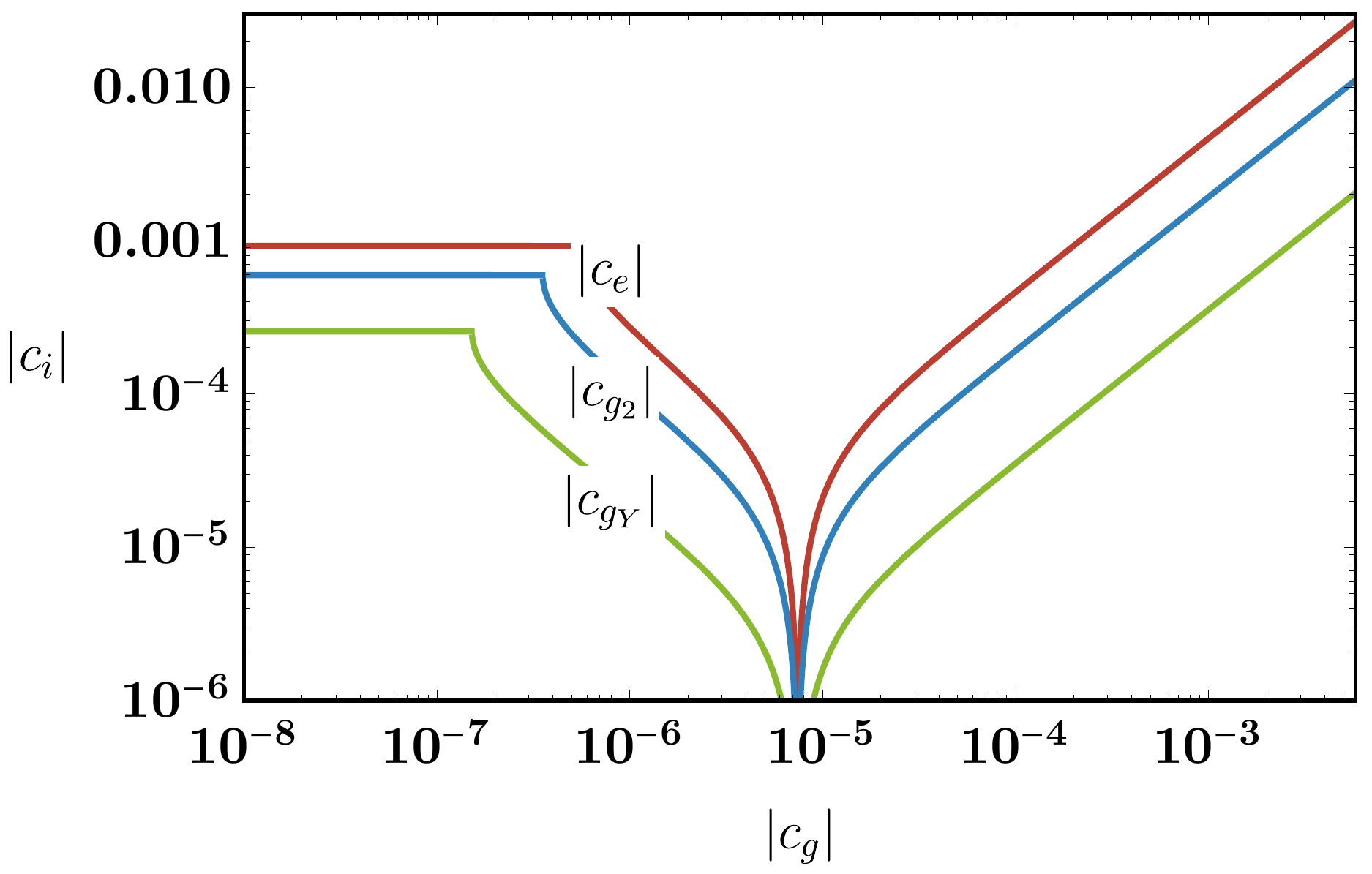} 
    \caption{Constraints from EP-violation searches from the E\"ot-Wash experiment \cite{Schlamminger:2007ht} on $c_e$, $c_{g_Y}$ and $c_{g_2}$ (blue, red, yellow) as a function of $|c_g|$, which is allowed to vary up to its maximum of $|c_g|\sim\sqrt{10^{-5}}$ as discussed in the text. The discontinuity at small values of $|c_g|$ corresponds to when the strongest constraint is on $\text{Im}(c_i)$ as opposed to $\text{Re}(c_i)$.}
    \label{fig:EPConstraints}
\end{figure}

There exist also constraints that can be set independently on $m_\p$ and $f_{\rm DM}$, coming from searches for EP-violating interactions. The strongest constraint was set by the E\"ot-Wash experiment \cite{Schlamminger:2007ht}, which compared the relative acceleration of Beryllium and Titanium test masses. Following the analysis of \cite{Damour:2010rp,Damour:2010rm}, we find that this constrains the combination of $c_e$ and $c_g$ to be
\begin{equation}
    \bigg| -4.2 \times 10^{-7} c_e^2 - 1.4\times10^{-3}c_e c_g + 6.6 \times10^{-3} c_g^2  \bigg| \lesssim (0.3 \pm 1.8) \times 10^{-13} \ .
\end{equation}
We use this constraint in \Cref{fig:EPConstraints} to show the upper limit on $c_e$, and the re-interpretation as an upper limit on $c_{g_Y}$ and $c_{g_2}$ in the absence of the other respectively. In the mass range of interest for $\p$, this, combined with the Cassini constraint on $c_g$, provide the strongest constraints.

Given these constraints, we see that the maximum scalar coupling to the $SU(2)_L$ kinetic term in the absence of tuning is $|c_{g_2}| \sim 10^{-2}$ if $c_g$ saturates the Cassini bound, or $|c_{g_2}| \sim 6 \times 10^{-4}$ if $|c_g|=0$. This constraint is so strong that a change in $\deltaW$ would not be possible, since the required scalar mass would be so small it would not be coherently oscillating now. These constraints might have been relaxed if a chameleon mechanism were present \cite{Khoury:2003aq}. However, given the requirement that the $\p^2$ term in the potential dominates over potential cubic and quartic terms, the Cassini constraint will still apply \cite{Blinov:2018vgc}.

Crucially however, these constraints are not directly on $\dgw$, but instead are interpretations of constraints on $c_e$ on the former. If we can impose that $c_e=0$, then there would be no existing constraint on $\dgw$ explicitly derived. This can be achieved if we impose a fine-tuning of the relation between $\dgy$ and $\dgw$, namely
\begin{equation}
    \frac{\dgy}{\dgw} = -\frac{\alpha_Y}{\alpha_2} \ ,
    \label{CouplingRel.EQ}
\end{equation}
which could have its origin in some symmetry. This relation corresponds to writing \Cref{ScalarCouplings.EQ} as
\beq
\Lag \supset -\frac{1}{4g_Y^2}B^{\mu \nu}{B}_{\mu \nu}-\frac{1}{4g_2^2}W^{a, \mu \nu}{W}^a_{\mu \nu}-\left(\frac{c_{g_2} \p}{g_2^2 \MPl}\right) \left( B^{\mu \nu}{B}_{\mu \nu} - W^{a, \mu \nu}{W}^a_{\mu \nu}\right) \ ,
\eeq
with $c_g$ being set to zero. The $B^2-W^2$ structure can arise, for example, from a left-right symmetric UV completion similar to that of \cite{Hook:2016mqo}, with $\p$ coupling to $W_R^2 - W_L^2$. This structure can be postulated at a given scale $\mu$, in which case a concern might be that unless it is enforced by an unbroken symmetry, it will not be invariant under RG flow. We will not consider this possibility here, and instead assume that the relation of \Cref{CouplingRel.EQ} holds in the IR.

In principle, in the limit $c_e=0$, a constraint on $\dgw$ can nevertheless be obtained by computing the EW matrix element of the proton and neutron. Variation of $\alpha_2$ would then lead to a small variation in the proton and neutron masses, which can be constrained by the E\"ot-Wash result. We expect this constraint to be roughly $\alpha_2/\alpha_{\text{EM}}\times \text{GeV}^2/m_W^2$ weaker than the constraint on $|c_e|$, which would translate into a bound $|\dgw| \lesssim 10\ (0.8)$ for $|c_g|=0\ (6\times10^{-2})$. A computation of the exact bound is left to future work.

\subsubsection{Scanning of $\alpha_2$}
Having established that $\dgw$ can be non-zero, and indeed could be sizeable if $c_e$ is set to zero by the tuning condition of \Cref{CouplingRel.EQ}, we now continue with the analysis of how $\alpha_2$ can be scanned by an ultra-light scalar.
\begin{figure}[t]
    \centering
    \includegraphics[width=0.49\textwidth]{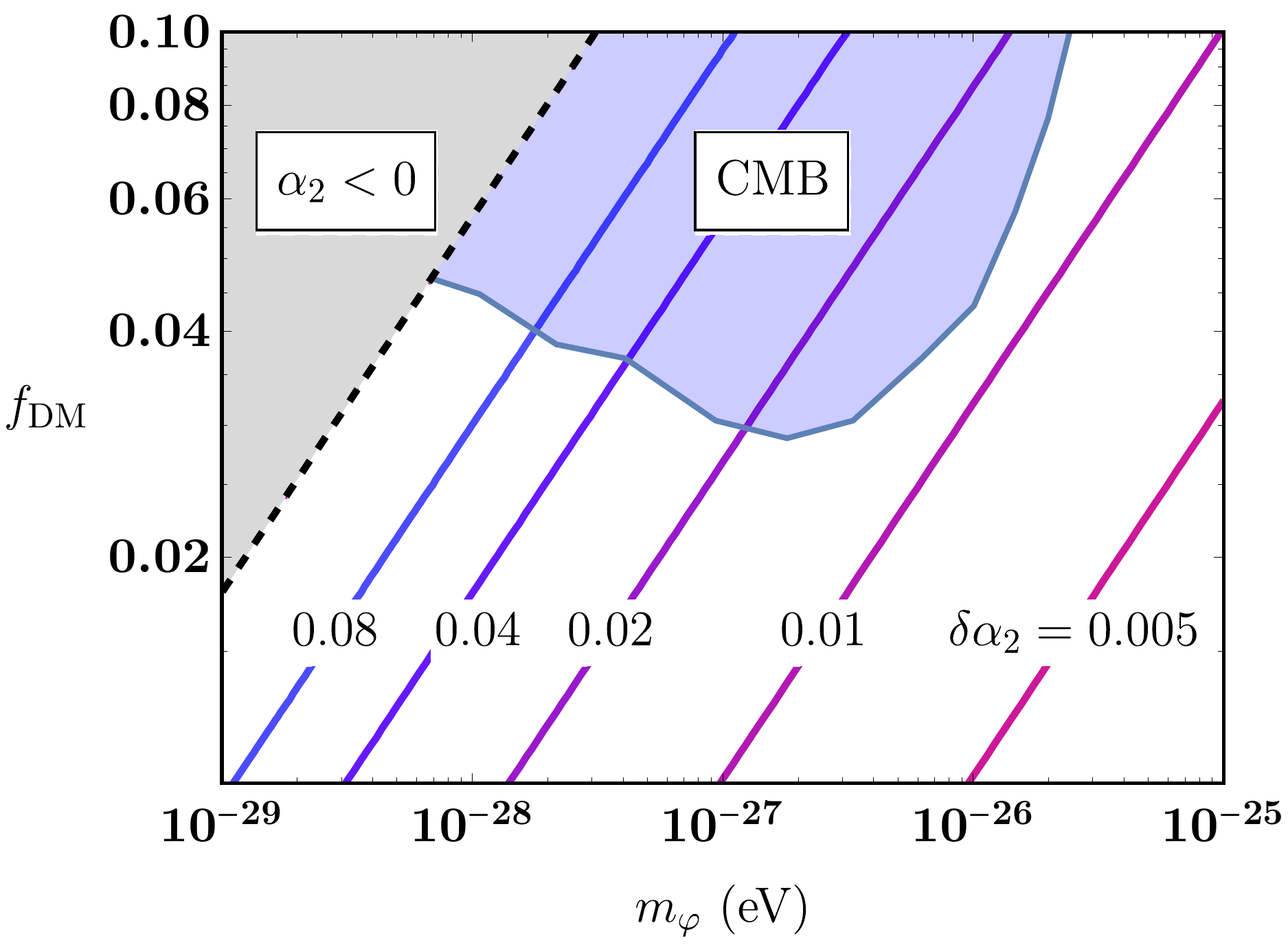}
    \includegraphics[width=0.49\textwidth]{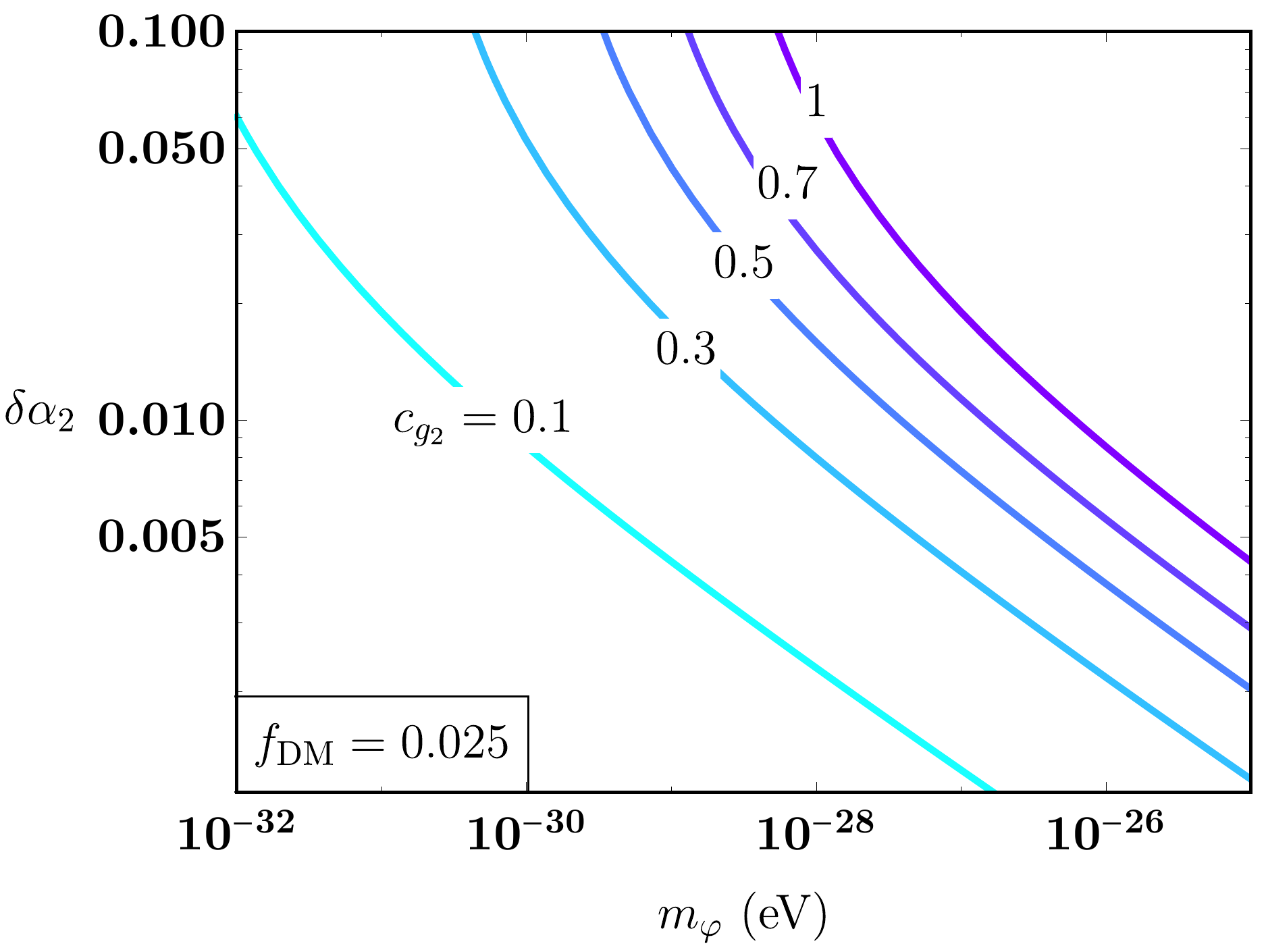}
    \caption{\textit{Left}: Contours of constant $\deltaW$ for varying scalar mass and DM fraction $f_{DM}$. The blue shaded region is excluded by CMB and large-scale structure (LSS) data \cite{Hlozek:2014lca}, while the dashed black line corresponds $\alpha_2 \to \infty$, with $\alpha_2 <0$ above. The scalar coupling has been set to $\dgw=1$. \textit{Right}: The variation in the weak coupling $\deltaW$ as a function of the mass for different choices of the scalar couplings $\dgw$. The DM fraction has been set to $f_{DM}=0.025$ to be free of CMB/LSS constraints. We allow large $\dgw$ by enforcing $c_e = 0$ as discussed in the text.}
    \label{fig:VarydW}
\end{figure}

The shift in the weak structure constant can be written as
\begin{equation}
    \deltaW = \alpha_2\frac{\dgw \p / \MPl}{1-\dgw \p / \MPl} \ ,
\end{equation}
with $\p$ given by \Cref{phi.EQ} such that 
\begin{equation}
    \deltaW \simeq \alpha_2 \left(\frac{\dgw \sqrt{f_{\rm DM}}}{2.5\times10^6\, \left(\frac{m_\p}{\text{eV}}\right)^{1/4}-\dgw \sqrt{f_{\rm DM}}} \right) \ ,
    \label{deltaWapprox.EQ}
\end{equation}
so that appreciable shifts in the weak structure constant will only occur for very small scalar masses, in the vicinity of the pole at $m_\p \sim 10^{-33}~{\rm eV}\,\left(\frac{\dgw}{0.1}\right)^4\, \left(\frac{ f_{DM}}{0.02}\right)^2$. This means that if $\p$ is to successfully scan $\alpha_2$ an appreciable amount, it must be ultra-light.

Given this expected range for the scalar mass, one must ensure that the ultra-light scalar is nevertheless sufficiently heavy that it is coherently oscillating at the present time. As discussed previously, this requires that the mass satisfies $m_\p \gtrsim 3 H_0 \sim 10^{-33}$ eV, where $H_0$ is the current value of the Hubble constant. Scalars satisfying this condition begin to coherently oscillate with an amplitude set by an initial misalignment in the field. In turn, this results in homogeneous energy densities that redshift with the scale factor $a(t)$ as $a(t)^{-3}$, just like non-relativistic matter. 

For masses below $m_\p \lesssim 10^{-20}$ eV, small-scale structure formation is suppressed on observable length scales \cite{Frieman:1995pm, Coble:1996te, Hu:2000ke}. However, a scalar with a light mass can still make up a fraction of the dark matter \cite{Amendola:2005ad, Hlozek:2014lca}. In the range of masses we will be interested in, the scalar may contribute to dark energy for some period of time after matter-radiation equality, before contributing to dark matter. This can change the heights of the acoustic peaks, meaning that measurements of the cosmic microwave background (CMB) are also strongly constraining in this regime. For masses between $10^{-32} \text{ eV} \lesssim m_\p \lesssim 10^{-26}$ eV, the scalar can only make up about 5\% of the dark matter energy density at 95\% C.L. \cite{Hlozek:2014lca}. 

Taking these constraints into account, we present our results in \Cref{fig:VarydW}. We see from the left plot that for a shift $\delta\alpha_2 \sim 0.01- 0.08$ as required to obtain the required baryon asymmetry (see \Cref{fig:EWBG}), if $\dgw = 1$, the scalar must have a mass $10^{-29.5} \text{ eV} \lesssim m_\p \lesssim 10^{-26.5}$ eV for $0.01 \lesssim f_{\rm DM} \lesssim 0.03$. The minimum fraction of dark matter the scalar could make up is bounded by the requirement $m_\p \gtrsim 3 H_0 \sim 10^{-33}$ eV, and is about $f_{\rm DM} \gtrsim 10^{-4}$. From the right plot, we see that relaxing the assumption that $\dgw=1$ but imposing that $f_{\rm DM} \sim 0.025$, the scalar must have a mass $10^{-32} \text{ eV} \lesssim m_\p \lesssim 10^{-26}$ eV. If ultimately it turned out that the E\"ot-Wash bound on $\dgw$ is $\mathcal{O}(10)$ as might be expected based on the discussion above, this mass range could be extended up to $m_\p \lesssim 10^{-19}$ eV, since the allowed fraction of dark matter would increase as well. Thus we find that our mechanism of scanning the weak coupling constant to be viable for a wide range of ultra-light scalar masses, and a variety of fractions of DM.

\subsection{Symmetry Breaking}

It has been shown that gauge couplings can be altered in the early universe via scalar fields that acquire a vev~\cite{Ipek:2018lhm}. Following this approach, we interpret $\p_{2,3}$ in \Cref{eq:model} as two real scalar fields, that are in a symmetric phase, with $\langle \p_i \rangle =0$, at high temperatures. The fields go through a symmetry breaking transition at a temperature $T_{\rm sb}\gtrsim 100~\GeV$, at which point they gain a non-zero vev $\langle\p_i\rangle$. (The two symmetry-breaking scales and the vevs might be generated through separate mechanisms and does not necessarily have a common source.) Due to this contribution, the gauge couplings are different than their SM values at the EW transition temperature. 

\begin{figure}[t]
    \centering
    \includegraphics[width=0.5\textwidth]{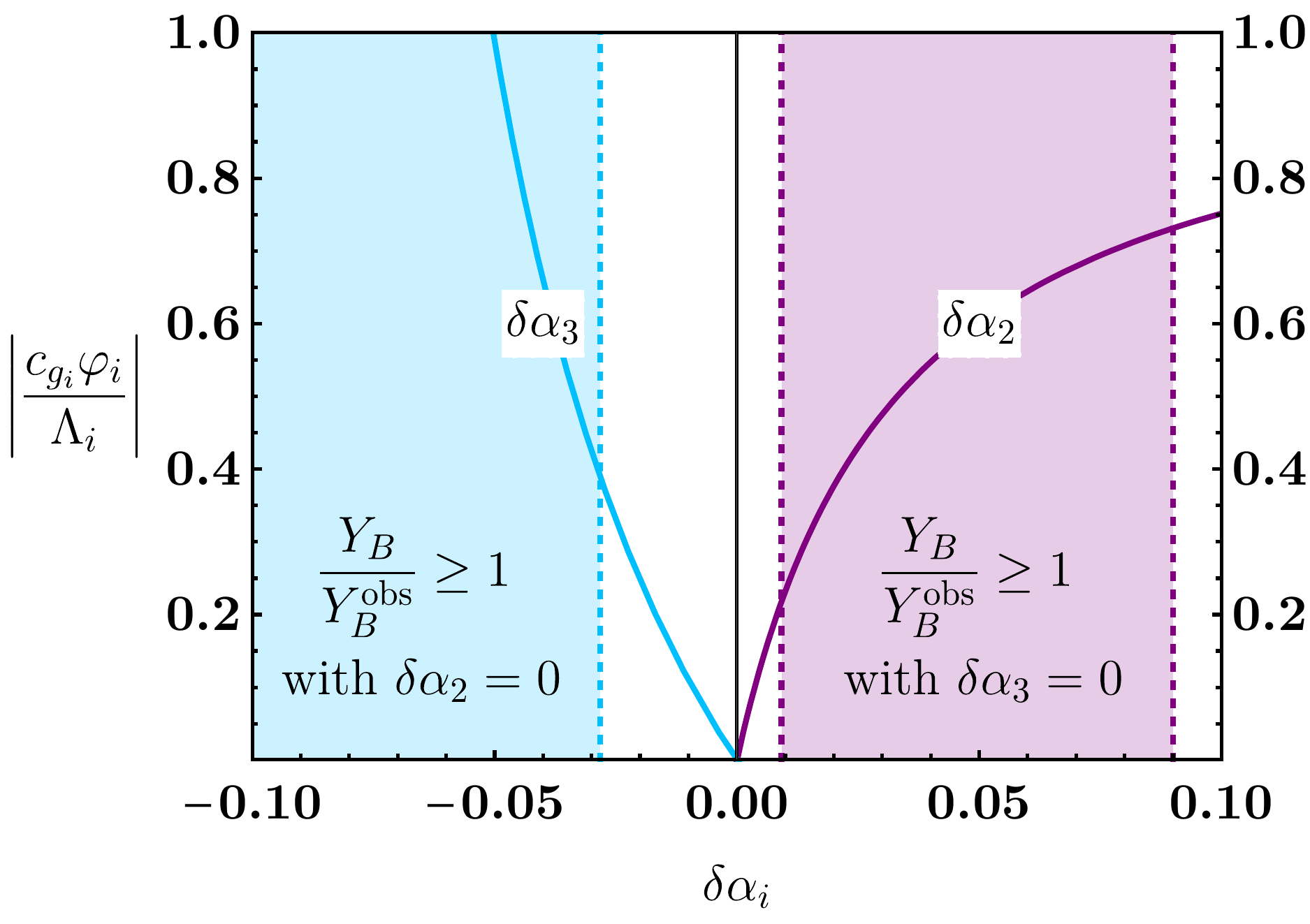} 
    \caption{The absolute value of the dimension-5 operators in \Cref{eq:model} for a given change in $\alpha_{2,3}$. In this scenario we assume the scalar field $\p_{2,3}$ acquires a vev $\langle \p_{2,3}\rangle$ before the EW phase transition, shifting the gauge couplings by $\delta\alpha_{2,3}$. In the purple shaded (on the right half of the plot), the BAU is produced without any change in $\alpha_3$, while in the blue shaded region (on the left half of the plot) BAU is produced without changing $\alpha_2$. Note that the produced asymmetry could be larger than observed in these regions.}
    \label{fig:dalphasymbreak}
\end{figure}

In the simplest scenario the dimension-5 terms in \Cref{eq:model} are generated by integrating out $N_f$ vector-like fermions in the representation $R$ of the gauge group, with mass $M_f$ and a Yukawa coupling $y_f$, charged under either $SU(2)_L$ or $SU(3)_c$. In this case we have
\begin{align}
\frac{c_{g_i}\langle\p_i\rangle}{\Lambda_i}\simeq N_f\, C(R)\,y_f\frac{2\alpha_i^2}{3}\frac{\langle\p_i\rangle}{M_f}~.
\end{align}
The invariant $C(R)$ is related to the quadratic Casimir operator, $C_2(R)$, through the relationship $C(R)=\frac{d_A}{d_R}C_2(R)$, where $d_A, d_R$ are the dimensions of the adjoint and $R$ respectively. For simplicity we turn on the interactions in \Cref{eq:model} one at a time, keeping one gauge coupling at its SM value while we change the other one. (See \Cref{fig:dalphasymbreak}.)

\textbf{Changing $\alpha_2$:} As discussed in \Cref{sec:BAU}, successful baryogenesis in the SMEFT scenario we consider can be achieved with a larger than expected $\alpha_2$ at $T_{c}\simeq 100~\GeV$. As can be seen in \Cref{fig:EWBG}, for $\delta\alpha_3=0$, successful baryogenesis requires $\delta\alpha_2 \gtrsim 0.01$. We can see from \Cref{fig:dalphasymbreak} that this requires $\left|\frac{c_{g_2}\langle\p_2\rangle}{\Lambda_2}\right|\simeq 0.2$. Taking $y_f\simeq 1,~ \left|\frac{\langle\p_2\rangle}{M_f}\right|\simeq 1$, this change in $\alpha_2$ requires $N_f\,C(R) \gtrsim 270$ fermions charged under $SU(2)_L$ with $M_f\sim O(\TeV)$.  Although it has been argued that $SU(2)_L$ has an asymptotically safe fixed point with such a large number of weak-scale fermions, the UV behavior of this scenario is not well understood. For example, it is expected that the theory be confined at very high scales, without a clear expectation to go into an unconfined phase below the Planck scale. Thus, within this scenario, we do not let the weak coupling to change from its SM value.

We also note that $\alpha_2$ can be altered via the RG running, by adding fermions charged solely under $SU(2)_L$. However, the required number of new charged fermions is vastly more than allowed by constraints on the EW precision parameter $W$, which was constrained by LEP to be smaller than $\Delta W \lesssim 8 \times 10^{-4}$. Additionally, strong LHC constraints can be placed on new $SU(2)_L$ charged fermions through modifications to the Drell-Yan process \cite{Alves:2014cda}.

\textbf{Changing $\alpha_3$:}  On the other hand, the BAU can also be generated by decreasing $\alpha_3$ during the EW phase transition.  In this case, with $\delta\alpha_2=0$, one needs $\delta\alpha_3\simeq -0.03$, which is achieved for $\left|\frac{c_{g_3}\langle\p_3\rangle}{\Lambda_3}\right|\simeq 0.4$. For $y_f\simeq 1,~ \left|\frac{\langle\p_3\rangle}{M_f}\right|\simeq 1$, and $N_f\,C(R)\simeq 60$, much less than the number of fermions required to change $\alpha_2$ by an amount that would generate the BAU. Such new fermions would likely be vector-like, in which case strong limits of $M_f\gtrsim \mathcal{O}$(TeV) apply from searches at the LHC if they have either $U(1)_Y$ or $SU(2)_L$ charges (see e.g. \cite{Aaboud:2018pii, Aaboud:2018ifs, Sirunyan:2019tib, Sirunyan:2018qau}).

\begin{figure}[h!]
    \centering
    \includegraphics[width=0.7\textwidth]{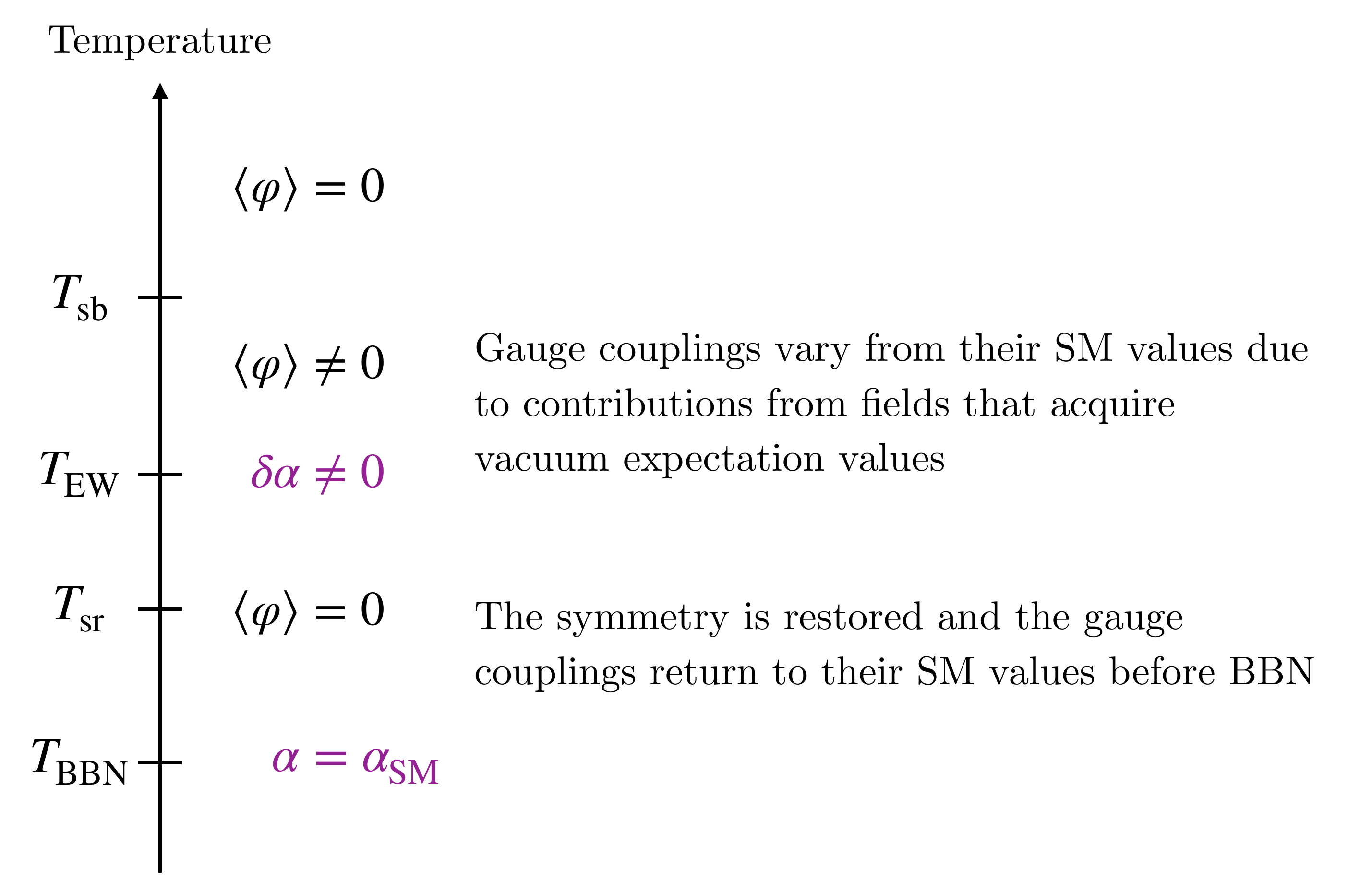} 
    \caption{An example of the cosmological history that is expected in the symmetry-breaking scenario.}
    \label{fig:scales}
\end{figure}

The couplings must return to their SM values sometime before BBN since the formation of light nuclei is well described within the SM. This can be achieved by another transition that restores the symmetry that was broken by $\langle \p_i\rangle$ at a temperature $T_{\rm sr}<T_{sb}$. We assume this symmetry restoration happens before BBN, $T_{\rm sr}\gtrsim \MeV$. (We will be agnostic as to whether these transitions are first order or not.) Thus, although the gauge couplings might differ from their SM values for a period of time in the early universe, we recover the SM before BBN. Such a symmetry breaking pattern can be present if there are multiple scalar fields which go under symmetry-breaking and restoring phases. (Similar symmetry breaking patterns have been studied in the literature, see \emph{e.g.}~\cite{Baker:2016xzo, Baker:2017zwx}.) A more detailed work on such a scenario is in progress~\cite{Ipeknew}.

\section{Conclusions}
\label{sec:conclusions}
We study the effects of varying weak and strong couplings in the early universe on the production of the observed baryon asymmetry. We do this in the context of the Standard Model with additional non-renormalisable operators, with the Lagrangian given in \Cref{eq:model}. We show that by raising the weak coupling constant and/or lowering the strong coupling constant around the EW phase transition scale we can easily produce the required baryon asymmetry. We do this in a model that was previously ruled out for baryogenesis purposes to demonstrate the power of these variations. The main reason for this success is that a larger weak coupling constant raises the weak sphaleron rate, which in turn produces a larger baryon asymmetry, despite the risk of increased washout by the same sphalerons in the broken phase of EW symmetry. Hence, the required extra \emph{CP} violation can be smaller compared to more traditional models. Similarly, lowering the strong coupling constant weakens the strong sphalerons, preventing the washout of a chiral symmetry, which, again, raises the produced baryon asymmetry. These effects can be seen in \Cref{CPV.FIG,fig:EWBG,fig:EWBGalpha3}. The viable parameter space can be almost entirely probed at HL-LHC, and entirely covered at proposed future lepton colliders.

We identify two types of models that would generate deviations in weak and strong coupling constants in the early universe. In one of these models, a light scalar couples to the $SU(2)_L$ and $U(1)_Y$ field strengths. This scalar field could constitute part of the dark matter and the couplings are constrained by various astrophysical and fifth-force experiments. Another model relies on a scalar field that undergoes symmetry breaking and symmetry restoration phases in the early universe. This model requires $O(50-300)$ fermions with $O({\rm TeV})$ masses, that are charged under $SU(3)_c$ or $SU(2)_L$ and is expected to have collider signatures.

\section*{Acknowledgements}
We thank Asher Berlin for helpful discussions. We thank David Morrissey for his thorough comments on a draft of this manuscript.
SARE is supported in part by the U.S. Department of Energy under Contract No. DE-AC02-76SF00515, and in part by the Swiss National Science Foundation (SNF) project P2SKP2\_171767.
SI acknowledges support from the University Office of the President via a UC Presidential Postdoctoral fellowship and partial support from NSF Grant No.~PHY-1620638. TRIUMF receives federal funding via  a  contribution  agreement  with  the  National  Research  Council  of  Canada  and  the  Natural Science and Engineering Research Council of Canada. This work was initiated and partly performed at the Aspen Center for Physics, which is supported by NSF grant PHY-1607611. We thank the participants of the workshop ``Understanding the Origin of the Baryon Asymmetry of the Universe" for excellent talks and discussions.
\newline
\newline
While this work was in completion, \cite{Danielsson:2019ftq} appeared on arXiv, which has overlap with \Cref{sec:scanningcouplings}.
\appendix

\section{Gravitational Wave Signals with a Varying Weak Coupling Constant}
\label{sec:GW}

It is well known that a strong first-order EW phase transition in the early universe produces gravitational waves (GWs) that are potentially visible at LISA \cite{Mazumdar:2018dfl,Caprini:2018mtu} and DECIGO \cite{Kawamura:2011zz}. In this section we study the GW spectrum produced in an EW phase transition, which is modified by both a $|H|^6$ term and by a change in $\delta\alpha_2$.

\begin{figure}
    \centering
    \includegraphics[width=0.7\textwidth]{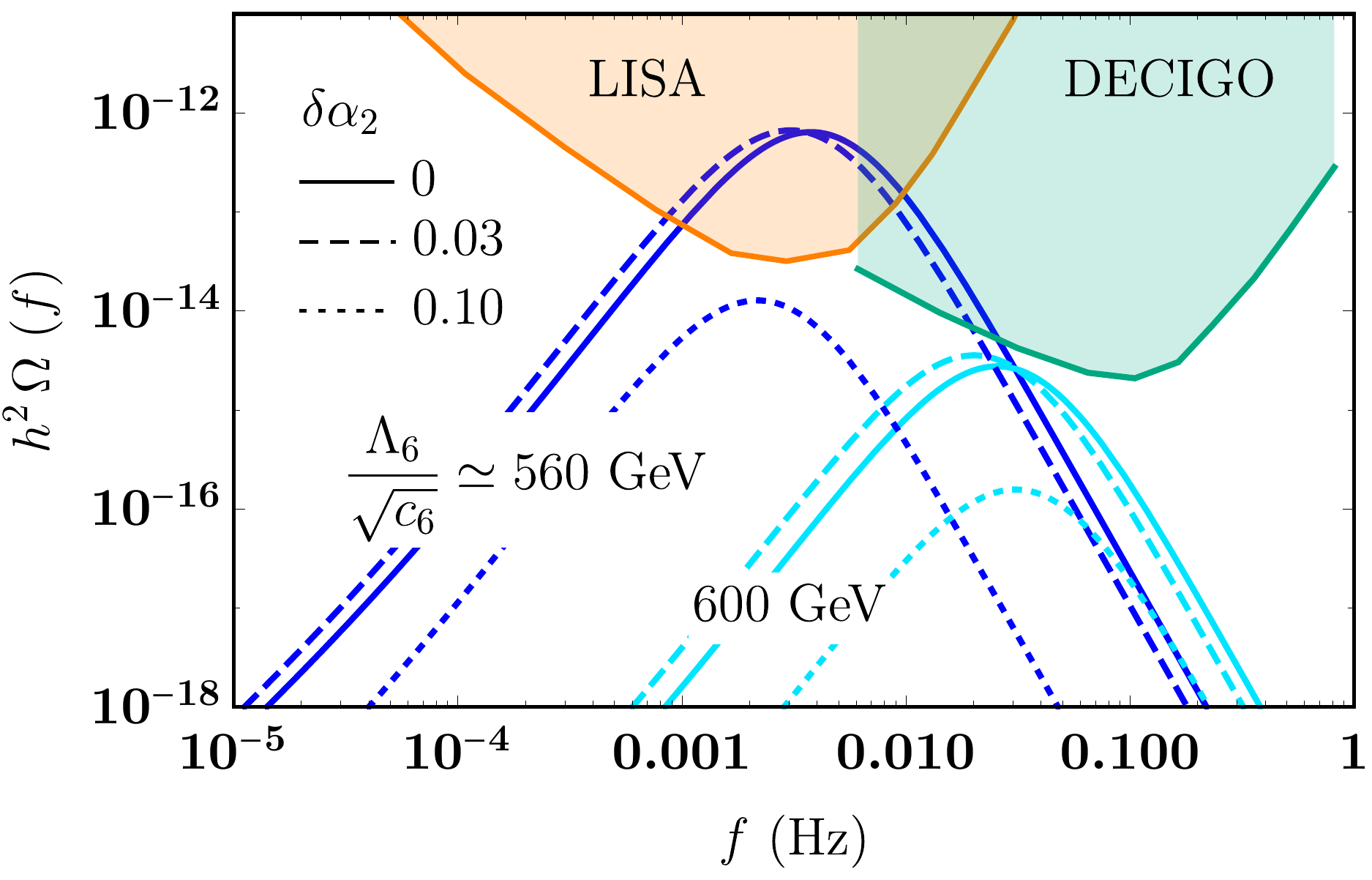}
    \caption{Gravitational wave power spectrum generated by $\Lambda_6/\sqrt{c_6}=560$~GeV (dark) and 600~GeV (light) and $\delta \alpha _2 = 0, 0.03, 0.1$ for $v_w\sim 0.5$. Note that $\delta\alpha_2=0.1$ does not give the observed BAU due to washout from weak sphalerons as discussed in \Cref{sec:BAU}.}
    \label{fig:GW_plot}
\end{figure}

The gravitational wave spectrum from a first-order cosmic phase transition includes three contributions \cite{Weir:2017wfa}
\begin{equation}
    \Omega (f) h^2 = \Omega _{\rm col} h^2 +\Omega _{\rm sw} h^2 +\Omega _{\rm turb} h^2 \ .
\end{equation}
Here the first term is generated via the collisions of bubbles. The soundwave contribution is due to the interactions of the bubbles with the plasma while the last term is due to turbulence. 
The sound wave contribution usually dominates when the Lorentz factor for the advancing bubble wall does not diverge \cite{Hindmarsh:2017gnf}. Indeed this is the regime we find ourselves in our scenario \cite{Bodeker:2009qy,Bodeker:2017cim,Croon:2018erz,Ellis:2018mja,Ellis:2019oqb}. The spectrum for the sound wave contribution is given by \cite{Hindmarsh:2013xza,Hindmarsh:2017gnf}
\begin{equation}\label{ampsw}
h^2\Omega _{\rm sw}    = 8.5 \times 10^{-6} \left( \frac{100}{g_*} \right)^{-1/3}   \left( \frac{\beta}{H} \right)^{-1} \Gamma ^2 \bar{U}_f^4 v_w S_{\rm sw}(f) 
\end{equation}
where $g_\ast$ is the number of degrees of freedom, $\Gamma \sim 4/3$ is the adiabatic index and $\beta/H$ describes the inverse duration of the phase transition with respect to the Hubble rate $H$. The rms fluid velocity is given by $\bar{U}_f^2\sim (3/4) \kappa _f \, \alpha $, where $\alpha$ defines the strength of the phase transition as the ratio of the latent heat and the entropy.  
We estimate the bubble wall velocity $v_w \approx 0.5 c$, for which the efficiency is well approximated by \cite{Espinosa:2010hh} 
\begin{equation}
    \kappa _f \sim \frac{\alpha ^{2/5} }{0.017+(0.997 + \alpha )^{2/5}}.
\end{equation}
The spectral shape is given by
\begin{equation} 
    S_{\rm sw} =  \left( \frac{f}{f_{\rm sw}} \right) ^3 \left( \frac{7}{4+3\left( \frac{f}{f_{\rm sw}}\right) ^2} \right)^{7/2}
\end{equation}
with the peak frequency
\begin{equation} \label{freqsw}
    f_{\rm sw} = 8.9 \times 10^{-8} \, {\rm Hz}\, \left(\frac{1}{v_w}\right) \left( \frac{\beta}{H} \right) \left( \frac{T_N}{{\rm GeV}} \right) \left( \frac{g_* }{100} \right)^{1/6} \left(\frac{z_p}{10} \right) \ ,
\end{equation}
where $T_N$ is the nucleation temperature and $z_p$ is a simulation factor which we take to be $6.9$~\cite{Hindmarsh:2017gnf}. Recent work has argued that for a SM Higgs potential modified by a higher dimensional $|H|^6$ operator, the sound waves do not last longer than the Hubble time \cite{Ellis:2018mja,Ellis:2019oqb}. This implies that simulations have overestimated the strength of the transition by a factor of
\begin{equation}
    v_w (8 \pi )^{1/3} \left( \frac{\beta}{H} \right)^{-1} \frac{4}{3\alpha \kappa _f} \sim O(10^2)\ .
\end{equation}
We present the GW spectrum with this suppression factor in the sound wave peak amplitude in \Cref{fig:GW_plot} for various values of $\delta \alpha _2$. Note that by including this suppression factor, we are presenting a pessimistic scenario. Recent simulations point to a more optimistic case.

The action for a critical bubble can be found using the public package \verb BUBBLE ~\verb PROFILER ~ 
\cite{Akula:2016gpl,Athron:2019nbd}. From the action as a function of temperature and $\deltaW$, we calculate the nucleation temperature and $\beta/ H$.  The general trend is that the peak amplitude reaches a maximal value for a moderate boost to $\delta \alpha _2$, beyond which the amplitude is suppressed. We conjecture the following explanation for this behavior. There is a competition between two effects when varying $\alpha_2$: \textbf{(i)} the thermal barrier is larger due to the contribution from a larger coupling and \textbf{(ii)} the Higgs potential evolves faster due to larger contributions to the thermal masses. The former effect enhances the GW signal while the latter suppresses it. This means there is an optimum value of $\alpha _2$ that maximizes the GW strength.

\section{Details of the Baryon Asymmetry Production}
\label{sec:appendix}
The calculation of the final baryon asymmetry can only be performed rigorously in toy models \cite{Cirigliano:2011di}. The approximate framework we use is known as the vev-insertion method. This  method has a number of assumptions that have never been properly tested, although there is ongoing work in this regard. These assumptions are as follows.
\begin{itemize}
    \item We can use the mass basis of the symmetric phase and ignore flavor oscillations. This is the assumption that most of the baryon asymmetry is produced in front of the wall where the vev is small and the mass matrix for quarks is basically diagonal with only the thermal masses.
    \item The dominant source of \emph{CP} violation comes from the collision term in the Boltzmann equations. We treat the semi-classical force as sub dominant. If the first assumption is valid then this second assumption appears to be also valid since this source of \emph{CP} violation tends to be  couple orders of magnitude larger than the semi classical force.
\end{itemize}
The baryon asymmetry is then calculated in three steps.
\begin{enumerate}
    \item  Calculate the bubble wall dynamics.
    \item  Use the output of step 1 to input into a set of transport dynamics which governs the behavior of number densities near a bubble wall. This will result in calculating a net chiral asymmetry that results from the \emph{CP}-violating source. Such a chiral asymmetry will be relaxed by strong sphalerons which wash out chiral asymmetries. It also acts as a seed to bias the weak sphalerons. That is, weak sphalerons convert the chiral asymmetry to a baryon asymmetry with an efficiency controlled by the weak sphaleron rate. 
    \item  Calculate EDM observables to make sure the \emph{CP}-violating parameters are not ruled out by experiments.
\end{enumerate}

\subsection{Step 1: Phase Transition}
First one needs to calculate the effective potential at finite temperature. This is achieved by calculating the one loop corrections to the effective potential using the finite temperature propagators. The effective potential can be written as a sum of zero temperature and finite temperature pieces,
\begin{equation}
    V(h,T)= V_0(h) + V_{\rm CW} (h) + V_T(h,T)
\end{equation}
where the first term is the tree level potential and middle term is the zero temperature, Coleman-Weinberg loop correction. Loop corrections will include all particles that interact with the Higgs. However, it is enough to focus on the ones with largest couplings, \emph{i.e.} Higgs self-coupling, gauge bosons and the top quark. There is also a question of gauge-dependence, as the calculations are necessarily done in a certain gauge choice. Physical quantities like the sphaleron energy is gauge-independent, while the measure of the order of the phase transition, namely $v_c/T_c$ is gauge-dependent. Although a gauge independent proxy has been developed~\cite{Patel:2011th}, it is not necessary to worry about this numerically.

The finite temperature corrections to the tree level potential are
\begin{equation}
    V_T =  \sum _{i \in {\rm bosons}}  n_i\frac{T^4}{2 \pi ^2} J_B\left[ \frac{m_i^2 + \Pi _i}{T^2} \right]+ \sum _{i \in {\rm fermions}}  n_i\frac{T^4}{2 \pi ^2} J_F\left[ \frac{m_i^2 }{T^2} \right] \ .
\end{equation}
In the above the $\Pi _i$ terms are Debye masses that result from a resummation and are included to prevent the breakdown of perturbation theory. See,\emph{e.g.}, \cite{Rose:2015lna}. Ignoring temperature corrections to the cosmological constant we can write a high temperature expansion for the thermal functions that are correct up to $m/T\lesssim 2$
\begin{equation}
    J_F (m^2/T^2) \sim -\frac{\pi ^2}{24} \frac{m^2}{T^2}\ , \quad J_B \sim \frac{\pi ^2}{12} \frac{m^2}{T^2} - \frac{\pi }{6} \frac{m^3}{T^3} \ . 
\end{equation}
We have dropped a log term which cancels the CW potential for a judicious value of the renormalization scale $\mu \sim T$.

At very high temperatures, the Higgs is in a symmetric phase with $v=0$. As the temperature drops, the potential acquires a second minima at non-zero vev. At a critical temperature $T_c$, these two minima become degenerate. The proxy for the strength of the phase transition is the ratio of the vev at $T_c$ to the critical temperature. The critical temperature is calculated by
\begin{equation}
    \left. \frac{dV}{dh } \right|_{h = v_c, T= T_c} =0, \quad V(v_c,T_c) = V(0,T_c) \ . 
\end{equation}
In order to obtain a large BAU, one requires $\frac{v_c}{T_c}>1$.
In the SM this is approximated as
\begin{equation}
    \frac{v_c}{T_c} \sim 0.1\, \frac{(g_2+\delta g_2)^3}{g_2^3} \ .
\end{equation}
In principle this suggests that a large $\delta g_2$ can catalyze a strongly first order phase transition. In practice, it is not possible to get larger than $v_c/T_c \sim 2/3$ for large values of $\delta g _2$.

Another way to get a strongly first order transition is by having a tree level barrier between the true and false vacuum that persists at zero temperature. This can be achieved by a single non-renormalizable operator added to the potential
\begin{equation}
    V_0 = \frac{\mu ^2}{2} h^2 -\frac{1}{4} \lambda_H h^4 +\frac{c_6}{8\Lambda_6^2}  h^6 \ .
\end{equation}
The alternating signs between the different powers of the Higgs fields allows for there to be a barrier between the true and false vacuum even at zero temperature.

The last quantity we need to calculate is the bubble wall profile. This comes from finding the bounce solution to the classical equations of motion
\begin{equation}
    h ^{\prime \prime }+\frac{2}{r} h ^\prime = \frac{dV}{d h}~ ,
\end{equation}
where the prime denotes a radial spatial derivative. The non-trivial solution to this equation is the bounce solution where one varies continuously from the true vacuum to the false one. The spatial profile of this solution approximates a $\tanh$ solution with three parameters: a bubble wall width $L_w$, the offset $\delta$ and the field value deep within the bubble $h _0$. We feed a numerical fit of the $\tanh$ profile into the transport equations described in the next section. This means the bubble wall parameters become functions of the temperature-varying couplings, \emph{i.e.} $L_w \to L_w(\delta \alpha_2 )$, etc. The bubble wall profile needs to be calculated at the nucleation temperature $T_N$. This temperature is defined when the action evaluated at the bounce satisfies
\begin{equation}
    \frac{S_E}{T_N} \sim 171 -4 \ln  \frac{T_N}{\rm GeV} -2 \ln g_* \ , \label{eqn:TN}
\end{equation}
where $g_\ast$ is the number of relativistic degrees of freedom. This condition implies there is at least one critical bubble in the Hubble volume. 

\subsection{Transport Equations}
In calculating the production of the baryon asymmetry, we split the calculation into two steps. First we calculate the baryon-conserving but \emph{CP}-violating interactions with the bubble wall including baryon- and \emph{CP}-conserving diffusion, scattering and decays. This allows one to calculate a profile for the total left-handed asymmetry. Then we feed this solution into the transport equation with weak sphalerons. Separating the calculation into two steps is justified on the grounds of the typical timescales involved in the first step $\tau _{\rm int} \sim 1/T $ is small compared to the time scale that governs baryon production $t_{\rm sph} \sim 1/ \Gamma _{\rm sph} $. Remarkably even when this is not true, the separation of the calculation into two steps only produces a few percent error. 

The transport equations that govern the first step given in the rest frame of the bubble wall for the right-handed top quark density $n_t$, the left-handed top doublet number density $n_Q$ and the Higgs density $n_H$ are
\begin{align}
    v_w n_t^\prime - D_t n_t^{\prime \prime } =&~ \Gamma _m \left( \frac{n_{Q}}{k_Q} - \frac{n_t}{k_t} \right)-\Gamma _Y \left( \frac{n_t}{k_t} - \frac{n_H}{k_H}- \frac{n_Q}{k_Q} \right) \\ 
    &~~~ + \Gamma _{\rm SS} \left( \frac{2 n_Q}{k_Q} - \frac{n_t}{k_t} +\frac{9(n_Q+n_t)}{k_B} \right) + S^{CPV}(z) \nonumber \\ 
    v_w n_Q^\prime - D_Q n_Q^{\prime \prime } = &~- \Gamma _m \left( \frac{n_{Q}}{k_Q} - \frac{n_t}{k_t} \right)+\Gamma _Y \left( \frac{n_t}{k_t} - \frac{n_H}{k_H}- \frac{n_Q}{k_Q} \right)\\
    &~~~ -2 \Gamma _{\rm ss} \left( \frac{2 n_Q}{k_Q} - \frac{n_t}{k_t} +\frac{9(n_Q+n_t)}{k_B} \right) - S^{CPV}(z) \nonumber \\ 
    v_w n_H^\prime - D_H n_H^{\prime \prime } =&~ \Gamma _Y \left( \frac{n_t}{k_t} - \frac{n_H}{k_H}- \frac{n_Q}{k_Q} \right)  
\end{align}
In the above we just guess the bubble wall velocity $v_w$ even though in reality it will depend on $\alpha_2 $ and $\alpha _3$.  We set $v_w=0.5$. The k factors are factors that relate the number densities to their chemical potentials. They turn out to be not important. The Yukawa term $\Gamma _Y$ is usually dominated by the ``scattering" contribution (decays tend to be kinematically suppressed) which involves a gluon and top annihilating into a Higgs boson and a top quark. This contribution is given by
\begin{equation}
    \Gamma _Y \sim 2 \times 0.129\, \frac{g_3^2 }{4 \pi}\, T \ .
\end{equation}
The strong sphaleron rate that washes out the chiral asymmetry is simply given by
\begin{equation}
    \Gamma _{ss} = 132 \alpha _3 ^5 T \ .
\end{equation}

The diffusion coefficients for the top quark are very sensitive to $\alpha_2$ and $\alpha _3$. They are given by \cite{Joyce:1994zn} (the discussion before Equation (130) in particular).
\begin{equation}
    D_Q^{-1} =   \epsilon _{L,R} \frac{45 }{7 \pi } \alpha_2 ^2 T \log \left( \frac{32 T^2}{M_W^2} \right) + Y_t^2 \frac{100}{7 \pi } \alpha_2^2 \tan ^4 \theta _w T \log \left( \frac{32 T^2}{M_B^2} \right)  + \frac{80}{ 7 \pi } \alpha _3^2  T \log \left( \frac{32 T^2 }{M_G^2} \right) \ , \label{eq:DQ}
\end{equation}
where $M_W = T \sqrt{20 \pi \alpha _2/3} $, $M_G =T \sqrt{8 \pi \alpha _3} $ and $M_B = T \sqrt{4 \pi \alpha _2 \tan ^2 \theta _2 /3} $.
Finally we have the interactions with the space time varying Higgs. Specifically these interactions are the left- and right-handed top quarks interacting with the vacuum. These are functions of the bubble wall profile (described in the previous section), the thermal widths through ${\cal E}_x = \omega _x - i \Gamma _x$ where $\Gamma _x$ is the thermal width, and the thermal masses. The \emph{CP}-violating interactions with the bubble wall are given by
\begin{align}
    S^{CPV} _{t}(z) =&~ \frac{3 v_W y_t ^2v_n^3(z)\partial _z v_n(z)}{\pi ^2} \nonumber \\ & \int \frac{k^2dk}{ \omega _L \omega _R} {\rm Im} \left[ ({\cal E}_L {\cal E}_R+k^2) \frac{n_F({\cal E}_L)+n_F({\cal E}_R)}{({\cal E}_L +{\cal E}_R)^2} + ({\cal E}_L {\cal E}_R^*-k^2)\frac{n_F({\cal E}_L)-n_F({\cal E}_R^*)}{({\cal E}_R-{\cal E}_L)^2} \right] 
\end{align}
where we have removed a divergent term by normal ordering \cite{Liu:2011jh}. Note that we neglect hole modes in the plasma. Due to this omission, we expect to underestimate the baryon asymmetry by a small amount \cite{Weldon:1999th,Tulin:2011wi}.  Similarly the relaxation term that describes \emph{CP}-conserving interactions with the bubble wall is given by
\begin{align}
    \Gamma _m = \frac{3y_t^2v^2_n}{4 \pi ^2 T}\int \frac{k^2dk}{\omega _L \omega _R}{\rm Im} \left[ ({\cal E}_L {\cal E}_R+k^2) \frac{h_F({\cal E}_L)+h_F({\cal E}_R)}{({\cal E}_L +{\cal E}_R)^2} + ({\cal E}_L {\cal E}_R^*-k^2)\frac{h_F({\cal E}_L)+h_F({\cal E}_R^*)}{({\cal E}_R-{\cal E}_L)^2} \right] \ . 
\end{align}
The relevant thermal widths are
\begin{equation}
    \Gamma _{L,R} = \frac{4}{3} \alpha _3 T \ .
\end{equation}
and the thermal masses are given by
\begin{align}
   m_{L}=& ~T\sqrt{\frac{g_3^2}{6}+\frac{3}{32} g_2^2 + \frac{1}{288} g_1^2 +\frac{1}{16} y_t^2 +\frac{1}{16}y_b^2} \\
   m_{R} =& ~T \sqrt{\frac{g_3^2}{6} + \frac{1}{18} g_1^2 + \frac{1}{8}y_t^2} \ . \label{eq:thermmass}
\end{align}
Both the \emph{CP}-conserving and \emph{CP}-violating interactions with the bubble wall have a resonant enhancement when $m_L \sim m_R$. The width and height of the resonance are controlled by the thermal width and the peak of the resonance is achieved when the thermal masses are degenerate. 

The second step in the calculation is in calculating the effect of weak sphalerons. For this we have a single transport equation
\begin{equation}
    D_Q \rho_B ^{\prime \prime } - \frac{15}{4} v_w \rho_B^\prime \Theta [-z] \Gamma _{\rm WS} = \Theta [-z] \frac{3}{2} \Gamma _{\rm WS} (5 n_Q(z) + 4n_t(z))
\end{equation}
where $\Gamma _{\rm sph} = 120 \alpha_2^5 T$. The baryon yield is then calculated by solving the above equation and dividing by the entropy density
\begin{equation}
    Y_B = \frac{3 \Gamma _{\rm WS}}{2 s D_Q \kappa _+}  \int _{-\infty} ^0 dy~ n_L (y) e^{-\kappa _- y}
\end{equation}
with $\kappa _\pm  = (v_w \pm \sqrt{v_w^2 +15 D_Q \Gamma _{WS} })/(2D_Q)$. To a very good approximation the left-handed number density in the symmetric phase can be approximated by a sum of exponentials, $n_L(y) \approx \sum _i A_i \exp B _i y$ \cite{White:2015bva}.  The baryon asymmetry is then
\begin{equation}
    Y_B = \frac{3 \Gamma _{\rm WS}}{2 s D_Q (v_w + \sqrt{v_w^2 +15 D_Q \Gamma _{ws} })/(2D_Q)} \sum _i \frac{A_i}{B _i - (v_w - \sqrt{v_w^2 +15 D_Q \Gamma _{\rm WS} })/(2D_Q)} \ .
\end{equation}

\bibliography{references}
\end{document}